\documentclass[12pt,twoside]{mitthesis}
\usepackage{lgrind}
\usepackage{cmap}
\usepackage[T1]{fontenc}
\usepackage{graphicx} 
\usepackage{hyperref}
\hypersetup{
    colorlinks=true,
    linkcolor=blue,
    filecolor=magenta,
    citecolor=SpringGreen4,
    urlcolor=cyan,
}
\usepackage{emptypage}
\usepackage{microtype}
\usepackage{caption}
\usepackage{subfig}
\usepackage{amsmath}
\usepackage{epigraph}

\usepackage{etoolbox}
\makeatletter
\newlength\epitextskip
\pretocmd{\@epitext}{\em}{}{}
\apptocmd{\@epitext}{\em}{}{}
\patchcmd{\epigraph}{\@epitext{#1}\\}{\@epitext{#1}\\[\epitextskip]}{}{}
\makeatother

\usepackage{pdfpages}
\usepackage{fixltx2e}
\usepackage[round]{natbib}
\usepackage[x11names]{xcolor}
\usepackage{braket}

\usepackage{array}
\newcolumntype{P}[1]{>{\centering\arraybackslash}p{#1}}
\usepackage{algorithm}
\usepackage{algpseudocode}
\usepackage{amsfonts}
\usepackage{epigraph}

\usepackage{titlesec}

\newcommand{\secfnt}{\fontsize{14}{17}}
\newcommand{\ssecfnt}{\fontsize{12}{14}}

\titleformat{\section}
{\normalfont\secfnt\bfseries}{\thesection}{1em}{}

\titleformat{\subsection}
{\normalfont\ssecfnt\bfseries}{\thesubsection}{1em}{}

\def\all{all}
\ifx\files\all  \else %\typeout{Including only \files.} \includeonly{\files} \fi

\usepackage{mathtools}
\begin{document}

% \renewcommand\cftchapfont{\normalsize\bfseries}
% \renewcommand\cftsecfont{\normalsize}

% \renewcommand\cftchappagefont{\normalsize}
% \renewcommand\cftsecpagefont{\normalsize}

% -*-latex-*-
% 
% For questions, comments, concerns or complaints:
% thesis@mit.edu
% 
%
% $Log: cover.tex,v $
% Revision 1.9  2019/08/06 14:18:15  cmalin
% Replaced sample content with non-specific text.
%
% Revision 1.8  2008/05/13 15:02:15  jdreed
% Degree month is June, not May.  Added note about prevdegrees.
% Arthur Smith's title updated
%
% Revision 1.7  2001/02/08 18:53:16  boojum
% changed some \newpages to \cleardoublepages
%
% Revision 1.6  1999/10/21 14:49:31  boojum
% changed comment referring to documentstyle
%
% Revision 1.5  1999/10/21 14:39:04  boojum
% *** empty log message ***
%
% Revision 1.4  1997/04/18  17:54:10  othomas
% added page numbers on abstract and cover, and made 1 abstract
% page the default rather than 2.  (anne hunter tells me this
% is the new institute standard.)
%
% Revision 1.4  1997/04/18  17:54:10  othomas
% added page numbers on abstract and cover, and made 1 abstract
% page the default rather than 2.  (anne hunter tells me this
% is the new institute standard.)
%
% Revision 1.3  93/05/17  17:06:29  starflt
% Added acknowledgements section (suggested by tompalka)
% 
% Revision 1.2  92/04/22  13:13:13  epeisach
% Fixes for 1991 course 6 requirements
% Phrase "and to grant others the right to do so" has been added to 
% permission clause
% Second copy of abstract is not counted as separate pages so numbering works
% out
% 
% Revision 1.1  92/04/22  13:08:20  epeisach
\title{Diffusion-Guided Renormalization of Neural Systems via Tensor Networks}

\author{Nathan X. Kodama}

\department{Department of Electrical, Computer, and Systems Engineering}

\degree{Doctor of Philosophy}

\degreemonth{May}
\degreeyear{2025}
\thesisdate{\today}

\maketitle

% First copy: start a new page, and save the page number.
\pagestyle{empty}
\cleardoublepage
% Uncomment the next line if you do NOT want a page number on your
% abstract and acknowledgments pages.
\pagestyle{empty}
\setcounter{savepage}{\thepage}
\begin{abstractpage}
% \input{abstract}
% \doublespacing
Far from equilibrium, fluctuation-driven neural systems self-organize across multiple scales towards efficient information processing and robust adaptations to external environments. Exploiting multiscale self-organization in systems neuroscience and artificial intelligence requires a computational framework targeted at modeling the effective non-equilibrium dynamics of stochastic neural trajectories. Non-equilibrium thermodynamics and representational geometry offer theoretical foundations for this framework, but we also need scalable data-driven techniques for modeling the collective properties of high-dimensional neural networks from partially subsampled observations.

Renormalization is a coarse-graining technique, which is central to the study of emergent scaling properties of many-body and nonlinear dynamical systems. While coarse-graining is widely applied to complex systems in physics and machine learning, coarse-graining complex dynamical networks is an unsolved problem affecting many computational sciences. The recent development of diffusion-based renormalization---inspired by quantum statistical mechanics---coarse-grains complex networks near entropy transitions marked by maximal changes in specific heat, or information transmission. Here I explore diffusion-based renormalization of dissipative neural systems by generating symmetry-breaking representations across multiple scales and offering scalable algorithms using tensor networks.

Diffusion-guided renormalization is the key innovation bridging microscale diffusion and mesoscale dynamics of dissipative neural systems. For microscale diffusion, I developed a scalable graph inference algorithm that discovers community structure from subsampled neural network activity. Using community-based node orderings, diffusion-guided renormalization efficiently models higher-order interactions and generate a renormalization group flow through metagraphs and joint probability functions. Towards mesoscales, diffusion-guided renormalization targets learning the effective non-equilibrium dynamics of dissipative neural trajectories, which occupy lower-dimensional subspaces of high-dimensional phase space. Ultimately, I extend diffusion-guided renormalization to coarse-to-fine prediction problems in systems neuroscience and artificial intelligence.

\end{abstractpage}

% Additional copy: start a new page, and reset the page number.  This way,
% the second copy of the abstract is not counted as separate pages.
% Uncomment the next 6 lines if you need two copies of the abstract
% page.
% \setcounter{page}{\thesavepage}
% \begin{abstractpage}
% \input{abstract}
% \end{abstractpage}

\cleardoublepage

% \section*{Acknowledgments}

%%%%%%%%%%%%%%%%%%%%%%%%%%%%%%%%%%%%%%%%%%%%%%%%%%%%%%%%%%%%%%%%%%%%%%
% -*-latex-*-

% Some departments (e.g. 5) require an additional signature page.  See
% signature.tex for more information and uncomment the following line if
% applicable.
% \include{signature}
\pagestyle{plain}
\raggedbottom
  % -*- Mode:TeX -*-
%% This file simply contains the commands that actually generate the table of
%% contents and lists of figures and tables.  You can omit any or all of
%% these files by simply taking out the appropriate command.  For more
%% information on these files, see appendix C.3.3 of the LaTeX manual. 
\pagenumbering{roman}
\tableofcontents
% \newpage
% \listoftables
% \newpage
% \listoffigures
% \newpage

\pagenumbering{arabic}
\chapter*{Preface}
\addcontentsline{toc}{chapter}{Preface}
Emerging computing paradigms at the intersection of physics and artificial intelligence have set the stage for a revolution in how we model and understand complex large-scale systems. Quantum machine learning~\citep{Biamonte2019} and thermodynamic AI~\citep{Coles2023} are reshaping the boundaries of what is computationally feasible and conceptually achievable, offering promising new frameworks to tackle problems previously thought intractable. These emerging methodologies emphasize principles from non-equilibrium thermodynamics, harness quantum mechanics' unique computational capabilities, and draw inspiration from quantum phenomena to enhance classical computation.

This dissertation sits at this dynamic intersection, motivated by the profound potential these paradigms hold for complex neural systems. Large-scale neural recordings offer vast repositories of data for systems neuroscientists investigating the functional relationships between fluctuation-driven brain activity and behavior~\citep{Urai2022}: Recently, self-supervised learning and disentangled representation learning techniques have been combined to discover latent dynamical embeddings of neural systems that are interpretable through the lens of behavior~\citep{Schneider2023,Liu2022,Batty2019}. 

Far from equilibrium, fluctuation-driven neural systems self-organize across multiple spatiotemporal scales toward efficient information processing and robust adaptations to external environments. Exploiting multiscale self-organization in artificial intelligence and systems neuroscience requires a computational framework targeted at modeling the effective dynamics of stochastic neural trajectories. Non-equilibrium thermodynamics and representational geometry offer theoretical foundations for this framework, but we also need fast algorithms and scalable data-driven techniques for inferring the collective properties of high-dimensional neural networks from partial observations.

\section*{Problem Statement}
\addcontentsline{toc}{section}{Problem Statement}
Inferring the collective properties of large-scale systems from partial observations is a major obstacle called the subsampling problem~\citep{Levina2022}.
One promising approach leverages renormalization group techniques from statistical physics to identify scale-invariant properties by iteratively coarse-graining subsampled neural networks~\citep{Meshulam2019}. 
% A promising approach to overcoming the subsampling problem is to find scale-invariant properties of neural systems by applying coarse-graining, renormalization group techniques from statistical mechanics~\citep{Meshulam2019}. 
Renormalization is central to the study of emergent scaling properties of many-body and nonlinear dynamical systems~\citep{Sethna2021,Kadanoff2013}. Although coarse-graining is widely applied to complex systems in physics and machine learning, coarse-graining complex networks is a challenging problem that affects many areas of computational science. 

\begin{figure}[h!]
    \centering
	\includegraphics[width=0.75\textwidth]{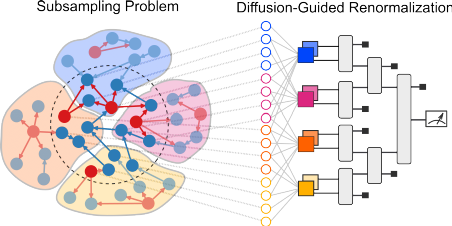}
     \caption{Subsampled neural networks may be coarse-grained through diffusion-guided renormalization, which identifies higher-order structures in latent graph communities.}
 \label{fig:subsampled-coarse-graining}
\end{figure}

Diffusion-based Laplacian renormalization is a recent proposal that coarse-grains complex networks near entropy transitions, marked by maximal changes in specific heat/information network diffusion in the network~\citep{Villegas2023}. It relies on a quantum-inspired framework for quantifying network information using spectral entropies~\citep{Domenico2016}. Here, a computationally sustainable alternative, called diffusion-guided renormalization, is developed to map disordered data from large-scale neural recordings---i.e., the irregular geometric domains of neural datasets---to multiscale tensor network renormalization techniques.

\section*{Strategic Overview}
\begin{figure}[h!]
    \centering
	\includegraphics[width=0.9\textwidth]{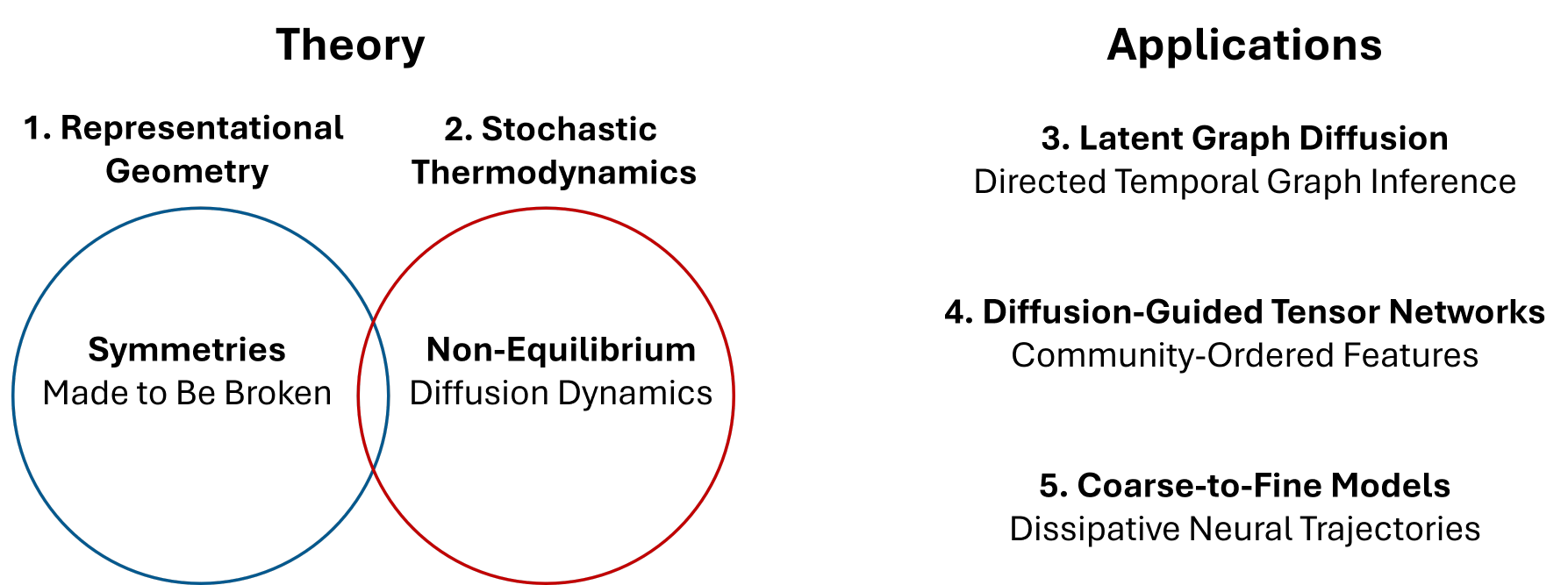}
     \caption{Strategy for inferring collective neural dynamics. Leveraging theoretical insights from the geometric thermodynamics of neural systems, I develop scalable algorithms for large-scale neural systems based on tensor networks, aimed at applications in systems neuroscience and artificial intelligence.}
 \label{fig:summary-strategy}
\end{figure}

\addcontentsline{toc}{section}{Strategic Overview}
In this dissertation, I introduce an interdisciplinary approach that integrates emerging computing paradigms---such as thermodynamic AI, quantum machine learning, and quantum-inspired algorithms---to address the challenge of modeling large-scale neural systems using generative learning. Specifically, I develop a multi-part strategy that leverages theoretical insights from stochastic thermodynamics and representational geometry to design scalable algorithms for inferring collective neural dynamics from large-scale neural recording data (Fig.~\ref{fig:summary-strategy}).

Geometric constraints---translational symmetry, permutation symmetries, and scaling properties---serve as informative priors that guide the design of interpretable and scalable algorithms. Stochastic thermodynamics offers mathematical descriptions of non-equilibrium neural systems. Asymmetric neural networks are modeled by diffusion processes across multiple spatiotemporal scales. By representing these diffusion processes through spectral analysis, we reveal distinct diffusion modes governing information flows and adaptive computations in neural systems. Representational geometry complements the thermodynamic viewpoint by providing a mathematical framework for handling symmetry-breaking transformations across multiscale neural representations.

\newpage
Based on these theoretical insights, I develop a scalable algorithm called Latent Graph Diffusion (LGD), which is aimed at efficiently inferring directed temporal graphs from subsampled neural data and simulating graph diffusion dynamics, uncovering coarse-grained network diffusion modes crucial for generating higher-order neural representations. These network diffusion modes are processed by expressive tensor networks---computational tools originally inspired by quantum many-body physics---which facilitate iterative coarse-graining of neural systems. Diffusion Tensor Network Renormalization (DTNR) generates multiscale renormalization group flows in the space of joint probability functions, enabling the inference of collective properties of neural systems.

Ultimately, diffusion-guided renormalization is applied to practical challenges at the interface of systems neuroscience and artificial intelligence. Coarse-to-fine predictive modeling frameworks are proposed in order to capture the effective non-equilibrium dynamics of neural trajectories. Such models are particularly suited for disentangling latent dynamical structures from large-scale neural recordings, enabling deeper insight into the neural mechanisms underlying behavior, cognition, and learning.

\subsection*{Geometric Thermodynamics of Neural Trajectories}
Non-equilibrium neural systems operate as open systems in continuous exchange with external environments and self-organize multiscale fluctuations to support adaptive learning and information processing. The rich dynamics observed in non-equilibrium neural systems can be effectively described through the lens of \textbf{stochastic thermodynamics}, particularly by focusing on the diffusion processes that characterize neural dynamics across multiple spatiotemporal scales. In this dissertation, these principles are leveraged to construct a theoretical and computational framework for analyzing neural systems, drawing insights from statistical physics to reveal the fundamental structures underlying non-equilibrium neural dynamics.

\begin{figure}[t!]
    \centering
	\includegraphics[width=0.9\textwidth]{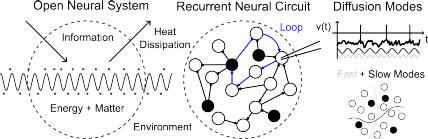}
     \caption{Open neural systems exchange energy, matter, and heat with the external environment. Recurrent neural networks give rise to stochastic diffusion modes across space and time. Diffusion modes have fast and slow components as revealed by spectral analysis of time-series and graphs.}
 \label{fig:open-neural-systems}
\end{figure}

Diffusion processes in non-equilibrium neural systems are mathematically described as discrete-time Markov chains or continuous-time Langevin dynamics. In particular, microscopic Langevin diffusion encompasses both deterministic drift components and stochastic diffusion terms driven by fluctuations. Such representations naturally capture the complex interactions and multiscale correlations intrinsic to asymmetrically coupled neural networks. By examining spectral decompositions of these stochastic diffusion modes, one uncovers a hierarchy of dynamical patterns ranging from rapid, microscopic neuronal interactions to slower, collective mesoscale dynamics (Fig. ~\ref{fig:open-neural-systems}).

Crucially, these dynamical patterns are not only described by stochastic thermodynamics but also by geometric principles that govern information processing in neural systems. Representational geometry provides a complementary framework to stochastic thermodynamics by focusing on the symmetries of neural signal domains and transformations of neural representations across different scales and modalities. Geometric insight---particularly those related to symmetry-breaking---offer crucial constraints for efficiently modeling neural data, enabling the systematic reduction of complexity and dimensionality inherent in neural trajectory analyses.

\begin{figure}[h!]
    \centering
	\includegraphics[width=0.9\textwidth]{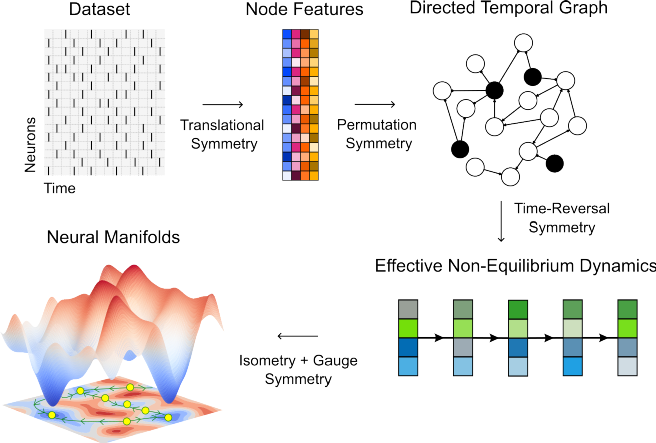}
     \caption{Representational geometry offers a unified approach to symmetry-breaking neural networks across multiple spatiotemporal scales. Translational symmetry of neuronal dynamics is used to build temporal node features. Permutation symmetry of unordered sets is constrained by the equivariance of directed temporal graphs. Time-reversal symmetry of effective non-equilibrium dynamics is broken by chaotic mixing and current flows. Isometry and gauge symmetries of neural manifolds are used to compress high-dimensional attractors into kinetic networks for neural codewords.}
 \label{fig:0-symmetries-made-broken}
\end{figure}

The combined insights from stochastic thermodynamics and representational geometry elucidate how neural systems evolve from microscale statistical neural mechanics to mesoscale collective neural dynamics~\ref{fig:0-symmetries-made-broken}. For example, in the process of inferring directed temporal graphs from observed neural data, the translational symmetry of single-neuron dynamics is preserved by temporal convolutions and the permutation symmetry of population-level neural interactions is preserved by graph convolutions. Toward mesoscales, however, inherent symmetries gradually break due to irreversible dynamics and coarse-graining procedures: effective non-equilibrium dynamics observed in dissipative neural trajectories break time-reversal symmetry, driven by chaotic mixing and entropic currents. Recovering geometric constraints requires one to turn to isometry and gauge symmetries of neural manifolds.

These complementary perspectives---the representational geometry of neural representations and the stochastic thermodynamics of neural trajectories---provide a unified theoretical approach to understanding multiscale non-equilibrium neural dynamics (Chapter~\ref{chapter:geometric-thermodynamics}). Using this unified approach, novel algorithms, such as latent graph diffusion and diffusion tensor network renormalization, are designed to effectively discover collective properties from large-scale neural recordings, facilitating predictive modeling, and advancing our understanding of neural information processing.

\subsection*{Diffusion-Guided Renormalization of Neural Systems}
To effectively analyze the multiscale dynamics of large-scale neural recordings, I introduce a computational framework for diffusion-guided renormalization. This computational framework is aimed at coarse-graining stochastic diffusion modes and generating multiscale renormalization group flows in the space of higher-order metagraphs or joint probability functions, enabling efficient models of collective neural dynamics from subsampled observations.

At the core of diffusion-guided renormalization is the latent graph diffusion (LGD) algorithm (Chapter~\ref{chapter:lgd}), which systematically transforms neural data into graph-structured representations. LGD involves three primary steps: (1) transformation of neural data into temporal node features, (2) simulation of graph diffusion processes that extract spectral diffusion modes, and (3) hierarchical coarse-graining of these diffusion modes into community-level features. Crucially, LGD leverages geometric priors---such as permutation equivariance of graph structures and translational equivariance in single-neuron dynamics---to reliably infer latent graph embeddings and guide subsequent coarse-graining.

Building on LGD's latent representation, diffusion tensor network renormalization (DTNR) generates higher-order representations of neural systems (Chapter~\ref{chapter:dtnr}). Hierarchical tensor networks efficiently generate a renormalization group flow in the space of joint probability functions. Using variational isometric compression techniques and feature maps inspired by coherent quantum feature learning, DTNR generates emergent multiscale structure, overcoming the curse of dimensionality and providing interpretable insights into the collective scaling properties of neural systems. By integrating LGD with tensor networks, DTNR identifies critical multiscale flows in total system entropy.

Together, LGD and DTNR offer a comprehensive tool for modeling complex neural systems. By generating stochastic and geometric representations across multiple scales, DTNR achieves a scalable, interpretable, and robust methodology for exploring multiscale phenomena, supporting predictive learning models and facilitating understanding of self-organization in non-equilibrium neural systems.

\subsection*{Coarse-to-Fine Models of Dissipative Neural Dynamics}
Dissipative neural trajectories occupy regional subspaces of a high-dimensional phase space by tracing out a strange attractor that is a highly non-trivial tangle of stable, neutral, and unstable manifolds~\citep{Engelken2020}. Discovering latent dynamical embeddings of dissipative neural trajectories is motivated by the prospect of learning interpretable mappings from neural activity to behavior~\citep{Schneider2023,Liu2022,Batty2019}. 
% Developing effective dynamical descriptions of dissipative neural trajectories is key to understanding information processing and governing laws of neural systems.

In Chapter~\ref{chapter:c2f}, I propose coarse-to-fine models that combine representation learning and tensor networks to systematically discover effective dynamics from dissipative neural trajectories. Learning the effective dynamics of dissipative neural trajectories with deep neural networks is a promising approach to discovering collective dynamical descriptions of neural systems~\citep{Vlachas2022,Pandarinath2018}. Tensor network variants of these models could exploit hierarchical structures generated by diffusion-guided renormalization algorithms. Crucially, the isometric transformations of tensor networks may efficiently learn parsimonious manifold embeddings of neural dynamics~\citep{Ehrlich2022,Chung2021,Cohen2020,Chaudhuri2019a,Chung2018}.

% Center manifolds of critically-stable neural dynamics are explored in Chapter~\ref{chapter:criticality}. 
% Dynamical modeling of neural recordings revealed that neural circuits display critically-stable dynamics, characterized by marginally stable eigenvalue spectra. 
Coarse-to-fine models capture relevant dynamical modes by representing neural states in a lower-dimensional space. Dynamical modes are closely related to spatial modes revealed by multiscale correlation analyses. Spontaneously active neural circuits in the neocortex self-organize into anti-correlated networks. These anti-correlated networks exhibit reciprocal fluctuations in activity, indicating structured interactions between distinct neural populations. Coarse-to-fine models may disentangle these anti-correlated networks by identifying distinct communities and revealing emergent interactions at larger length scales.

Overall, these applications underscore the potential of coarse-to-fine models for understanding neural computations across scales. By combining geometric thermodynamic and diffusion-guided renormalization, coarse-to-fine models are a powerful tool for discovering the effective dynamics of dissipative trajectories on neural manifolds.

% \begin{figure}[!h]
%   \centering
%   \includegraphics[scale=0.8]{templates/figures/proposal/fig1.pdf}
%   \caption{Coarse-to-fine models capture relevant spatiotemporal modes by representing neural states in a lower-dimensional space.}
%   \label{fig:overview-f1}
% \end{figure} 

\section*{Contributions}
\addcontentsline{toc}{section}{Contributions}
In this dissertation, a unified theoretical and computational framework that integrates emerging computing paradigms---thermodynamic AI, quantum machine learning, and quantum-inspired algorithms---to address pressing challenges in modeling large-scale recording data. Scalable algorithms are developed from theoretical insight and are connected to proposed applications aimed at developing our understanding of neural dynamics across multiple spatiotemporal scales. The main contributions of this dissertation are summarized as follows:

\begin{enumerate}
    \item \textbf{Unified Theoretical Framework}: Offered geometric thermodynamics of neural trajectories as a unified  theoretical perspective integrating stochastic thermodynamics with representational geometry to characterize symmetry-breaking transformations and emergent scaling behavior in neural systems.
    \item \textbf{Latent Graph Diffusion (LGD) Algorithm}: Developed a scalable algorithm for inferring directed temporal graphs and community-level features from neural data, facilitating higher-order neural representation.
    \item \textbf{Diffusion Tensor Network Renormalization (DTNR)}: Constructed an expressive tensor network framework that implements variational algorithms to iteratively coarse-grain neural systems, generating a multiscale renormalization group flow in the space of joint probability functions.
    \item \textbf{Coarse-to-Fine Modeling}: Proposed paths towards key applications, leveraging tensor networks and representation learning to discover the effective non-equilibrium dynamics of dissipative neural trajectories.
    \item \textbf{Applications to Neural Systems}: Presented  emergent phenomena---critically-stable dynamics and anti-correlated networks---serving as empirical evidence of spatiotemporal diffusion modes.
\end{enumerate}

These contributions advance the understanding, interpretability, and scalability of generative learning models in artificial intelligence and systems neuroscience, offering a robust framework for future developments at the intersection of physics and machine learning.
% \part{Theory}
\chapter{Geometric Priors for Neural Systems}\label{chapter:geometric-thermodynamics}
% \chapter{Geometric Thermodynamics of Diffusion in Neural Trajectories}\label{chapter:geometric-thermodynamics}

Unifying neural representations across different tasks, modalities, and scales is a high-dimensional problem that is best approached using \textit{a priori} knowledge about the representational geometry of neural networks~\citep{Ehrlich2022,Chung2021}. Geometric priors---i.e., first-principles of symmetry, scaling, and deformation stability from geometric deep learning~\citep{Bronstein2021} and statistical physics~\citep{Sethna2021}---are crucial to overcoming the curse of dimensionality. However, recent advances in %symmetry-preserving,
equivariant neural networks have revealed %practical problems associated with trainability and more 
practical challenges and potential solutions for discovering symmetry-breaking in physical systems across multiple length scales~\citep{Wang2024}.
In this chapter, geometric priors for learning neural representations are developed by focusing on the symmetry constraints of neural networks across scales.

Symmetries are preserved or broken in neural network, as neural data are transformed on spatial graphs and temporal grids or  coarse-grained on latent manifolds. 
Relevant domain symmetries of neural signals and emergent/broken symmetries of coarse-grained neural systems are explored. 
Furthermore, \textit{a priori} knowledge of the network connectivity of biophysical neural networks is used to further reduce the hypothesis space of learning representations. Cortical circuits, for example, are organized into sparse, sign-directed, and hierarchical modular communities. These symmetries and structural constraints reduce the hypothesis space for representation learning models.

% In both many-body physics and deep learning, the symmetry of domains underlying neural signals and transformations across multiple scales are fundamental. The implications of several key symmetries are explored analytically and numerically: permutation symmetry of graphs, translational symmetry of grids, time-reversal symmetry, and scale-invariance. Preserved and broken symmetries could serve as the basis for potential statistical field theories for neural systems, which is a direction that we are working towards.

% \begin{figure}[t!]
%     \centering
% 	\includegraphics[width=0.9\textwidth]{templates/figures/symmetries-made-broken.pdf}
%      \caption{Representational geometry offers a unified approach to symmetry-breaking neural networks across multiple spatiotemporal scales. Translational symmetry of neuronal dynamics is used to build temporal node features. Permutation symmetry of unordered sets is constrained by the equivariance of directed temporal graphs. Time-reversal symmetry of effective non-equilibrium dynamics is broken by chaotic mixing and current flows. Isometry and gauge symmetries of neural manifolds are used to compress high-dimensional attractors into kinetic networks for neural codewords.}
%  \label{fig:2-symmetries-made-broken}
% \end{figure}

% \subsection{Effective dynamics compress neural information}
\section{Symmetries made to be broken by neural networks}~\label{sec:symmetry}
Representation geometry formalizes the symmetries of a \textit{domain} $\Omega$---e.g., a temporal grid, spatial graph, or dynamical manifold---in the language of group theory. Geometric machine learning with classical deep neural networks~\citep{Wang2024, Bronstein2021} and quantum variational circuits~\citep{Nguyen2022,Ragone2022,Nielsen2010} operate on \textit{signals} $x(u) \in \mathcal{X}(\Omega)$ on the domain of the form $f: \mathcal{X}(\Omega) \rightarrow \mathcal{Y}$ by discovering \textit{transformations} $f(x)$ in some \textit{hypothesis class} $\mathcal{F}(\mathcal{X}(\Omega))$. 

% \subsection{Exact symmetries: groups, actions, and representations}
Symmetries correspond to groups of transformations that leaves a set of quantities unaltered (\textit{invariant}) or similarly altered (\textit{equivariant}). For example, the energy of a spin-glass neural network is invariant to permutations of individual neurons, because the Hamiltonian sums over the state of each neuron. On the other hand, the state vector and coupling matrix are equivariant to permutations of individual neurons. Permutation of neurons $i$ and $j$ in the network is must be accompanied by a permutation of rows $i$ and $j$ in the state vector and rows/columns $i$ and $j$ coupling matrix to recover an equivalent dynamical description.

% \section{Broken symmetries of nonequilibrium neural networks} \label{sec:permutation}
% \subsection{Permutation on graphs, translation on grids, and isometries on manifolds} \label{sec:permutation}
For neural systems, there are a few key symmetries. Here we focus on the permutation symmetry of neural networks and translational symmetry in neural dynamics. Microscopic descriptions of neural systems take the form of complex networks of neurons with adaptive dynamics. Because neural signals are described by functions on some domain---e.g., grids, graphs, and manifolds---identification, transformation, and representation on the appropriate domain is key to obtaining effective dynamical models. 

Permutation symmetry is fundamental to analyzing discrete spatial representations of neural systems. Often, a canonical ordering of individual neurons is not given \textit{a priori}. As a result, neural populations (sets) and neural networks (graphs) exhibit invariant and equivariant properties under permutations of neurons. Exploiting preserved permutation symmetries in finding reduced representations of neural systems requires the construction of permutation invariant and permutation equivariant transformations. On the other hand, one might be interested in breaking global permutation symmetry by discovering canonical orderings of individual neurons according to the modular structure of neural systems: this requires the construction of relaxed permutation invariant and permutation equivariant transformations that are stable to local deformations.

Translational symmetry is fundamental to analyzing neural dynamics (temporal grids). At microscales, the steady-state dynamics of individual neurons may exhibit local translational symmetry. Convolutional kernels capture quasistatic synaptic interactions and autoregressive spike responses. Toward mesoscales, the global translational symmetry of neural dynamics is broken by long-range adaptations and spatiotemporal interactions. Time-reversal symmetries of dynamical systems, such as in Newton's second law or equilibrium thermodynamics under conditions of detailed balance, are key. For example, magnetic spin systems break time-reversal symmetry: reversing the arrow of time changes the sign of the magnetization. Determining whether such symmetries should be preserved or broken is a key research area.

% \subsection{Permutation symmetry of graphs}
\subsection{Permutation symmetry of neural networks}
Permutation symmetry is fundamental to analyzing discrete spatial representations of neural systems. Often, a canonical ordering of individual neurons is not given \textit{a priori}. As a result, neural populations (sets) and neural networks (graphs) exhibit invariant and equivariant properties under permutations of neurons. Exploiting preserved permutation symmetries in finding reduced representations of neural systems requires the construction of permutation invariant and permutation equivariant transformations. 

Permutation symmetries arise naturally when characterizing systems that lack canonical orderings, such as neural populations (sets) and neural networks (graphs).  Consider a neural system constituted by $n$ neurons, the state of neuron $i$ is described by a $d$-dimensional vector, denoted by $\mathbf{x}_i$. The symmetric group $\mathcal{S}_n$ contains elements that give all possible orderings of the set of indices.
% where each neuron is equipped with a local $d$-dimensional basis $|i\rangle, i=1 \ldots d$ The symmetric group $\mathcal{S}_n$ contains elements that give all possible orderings of the set of site indices $\{1, \ldots, n\}$. There are exactly $n!$ permutations, so $\mathcal{S}_n$ is a very large group.
% $$
% \mathbf{P}(\pi)[i_1, i_2, \cdots, i_n]=[i_{\pi^{-1}(1)} i_{\pi^{-1}(2)} \cdots i_{\pi^{-1}(n)}]
% $$
% where $\pi \in \mathcal{S}_n$ and $[i_1 i_2 \ldots]$ is a shorthand for $[i_1] \otimes[i_2] \otimes \ldots$.
By stacking the neuron features as rows of the $n \times d$ matrix $\mathbf{X}=\left(\mathbf{x}_1, \ldots, \mathbf{x}_n\right)^{\top}$, an arbitrary ordering of the neurons has been imposed. The action of the permutation $\pi \in \mathcal{S}_n$ on the collection of neurons permutes the rows of $\mathbf{X}$, which can be represented as a $n \times n$ permutation matrix $\rho(\pi) = \mathbf{P}$.

\subsubsection{Global permutation invariants of neural populations}
A permutation invariant function $f$ acting on the state of the neural population (set) satisfies $f(\mathbf{P} \mathbf{X})=f(\mathbf{X})$  for all $\mathbf{P}$. One possible function is
$$
f\left(\sum_{v \in \mathcal{V}} s\left(\mathbf{x}_u\right)\right)
$$
where the function $s$ is independently applied to every node's features, and $f$ is a function of the sum over outputs. Because the sum is independent of the ordering,  $f$ is invariant under permutations. At macroscales, extensive thermodynamic properties and topological invariants such as the energy, entropy, free energy, total activation (magnetization), attractor dimension, and critical exponents are permutation invariant properties of neural populations. Interestingly, there is a privileged basis for mesoscale descriptions analogous to total angular momentum, which is given by irreducible representations via the Schur transform.

\subsubsection{Local permutation equivariants of neural networks}
A graph $\mathcal{G}=(V,E)$ contains \textit{nodes} $V$ and edges $E \subseteq V \times V$ coupling two nodes. The state of each node is described by a $d$-dimensional node feature vector, denoted by $\mathbf{x}_v$ for all $v \in V$. Define the $d$-dimensional node-wise signals as $\mathcal{X}\left(\mathcal{G}, \mathbb{R}^d\right)$. Time-evolution updates of neural networks stack the global permutation invariant functions of neural subpopulations as 
$$
\mathbf{f}(\mathbf{X}, \mathbf{A})=\left[\begin{array}{ccc}
- & f\left(\mathbf{x}_1, \mathbf{X}_{\mathcal{N}_1}\right) & - \\
- & f\left(\mathbf{x}_2, \mathbf{X}_{\mathcal{N}_2}\right) & - \\
& \vdots & \\
- & f\left(\mathbf{x}_n, \mathbf{X}_{\mathcal{N}_n}\right) & -
\end{array}\right]
$$
where $\mathbf{A}$ is the \textit{adjacency matrix} defined as 
$$
a_{ij}= \begin{cases}1 & (i, j) \in E \\ 0 & \text { otherwise }\end{cases}
$$
and $\mathbf{X}_{\mathcal{N}_i}=\left\{\left\{\mathbf{x}_j: j \in \mathcal{N}_i\right\}\right\}$ is the multiset of neighborhood feature and $\mathcal{N}_i=\{j:(i, j) \in E\}$ is the neighborhood of node $i$.

A global permutation invariant function $f$ for the graph satisfies
$$
f\left(\mathbf{P X}, \mathbf{P A} \mathbf{P}^{\top}\right)=f(\mathbf{X}, \mathbf{A})
$$
and a local permutation equivariant function $\mathbf{f}$ satisfies
$$
\mathbf{F}\left(\mathbf{P X}, \mathbf{P A P}^{\top}\right)=\mathbf{P F}(\mathbf{X}, \mathbf{A})
$$
for any permutation matrix $\mathbf{P}$.
% \subsubsection{Approximate permutation invarance and equivariance}
From microscale to mesoscale thermodynamics, there are several permutation equivariant objects for Langevin diffusion neural networks and coarse-grained neural metagraphs, including the state vector, drift matrix, diffusion matrix, adjacency matrix, Laplacian matrix, and network propagator.

\subsection{Approximate symmetry with deformation stability}\label{sec:deformation}
% \subsection{Exact symmetry via group theory}
% Consider the possible domains underlying neural signals and neural systems---grids, graphs, and manifolds. Assume the machine learning system operates on signals that are functions on some domain $\Omega$, i.e. there is a domain underlying the data. Formally, a point on the domain is $u \in \Omega$ and the signal on the domain is $x(u) \in \mathcal{X}(\Omega,\mathcal{C})$, where $x$ is a function or mapping $x:\Omega \rightarrow \mathcal{C}$. The space of $\mathcal{C}$-valued signals on $\Omega$ is a function space that has a vector space structure. $\mathcal{C}$ is a function space whose dimensions are called \textit{channels}. Thus, $\mathcal{X}(\Omega, \mathcal{C}) = \{x: \Omega \rightarrow \mathcal{C}\}$. When $\Omega$ has additional structure, we may further restrict the kinds of signals in $\mathcal{X}(\Omega, \mathcal{C}).$ What are the relevant signal domains underlying neural signals?

%  of hybrid transformations
Although group theory gives formal mathematical language for symmetries, there are technical challenges arising in real-world, noisy data (Table~\ref{table:symmetry-constraints}). Global symmetries may be broken across multiple scales or critical transitions. \cite{Wang2024} offers an attractive approach to discovering broken translational symmetry with relaxed group convolutions with symmetry-breaking neural networks. This approach fits within a more general framework of relaxing symmetry constraints with deformation stability and multiscale representation~\citep{Bronstein2021}. Relaxing symmetry with deformation stability enables the discovery of approximate symmetries that are hidden or emergent, and therefore hard to define \textit{a priori}.

\begin{table}[h!]
\centering
\caption{Challenges for symmetry-constrained neural networks.}
\begin{tabular}{||c c||} 
 \hline
 \textbf{Challenge} & \textbf{Approach} \\ [0.5ex] 
 \hline\hline
 Broken global symmetry & Approximate local symmetry \\
 Discovering approximate symmetries & Minimal-complexity models\\
 Instability of momentum-space representation & Hierarchy of hybrid-space representation\\
 [1ex] 
 \hline
\end{tabular}
\label{table:symmetry-constraints}
\end{table}

Data augmentation techniques are commonly employed in learning algorithms to find models that are invariant or equivariant to certain transformations, e.g. translations and rotations. Alternatively, restricting the hypothesis space to minimal complexity models with invariant and equivariant weight-sharing constraints is more computationally efficient.  Balancing expressive power and model complexity is crucial to building trainable models with fast convergence and generalization. Successfully navigating the landscapes of minimal-complexity learning models is non-trivial and falls within the broader subject of meta learning. Discovering models that offer the best performance benefits from well-defined complexity measures. From the perspective of Bayesian scientific computing, these complexity measures guide the selection of geometric priors. Deformation stability offers an analytical framework for defining complexity measures with respect to global symmetry groups.

\section{Emergent scaling properties of multiscale neural representations}
Emergent scaling structure is another geometric prior that is essential to learning representations of symmetry-breaking neural networks. In statistical physics, emergent properties are revealed by the renormalization group: a collection of coarse-graining transformations that generate a flow of multiscale models. Studying the renormalization group flow enables the study 
of scaling properties of critical and dynamical systems. The deformation stability of neural signal domains is briefly introduced because it determines choices between different classes of transformations and representations. Crucially, partial knowledge of geometric structures underlying neural signals (e.g., network coupling) can conspire with misaligned domain transformations
% (e.g., random vs. windowed sampling) 
to bias estimates of collective properties and thwart system identification~\citep{Levina2022}, so priors neural coupling and fluctuations must be applied. Emergent scaling of neural representations and deformation stability of domain transformations are explored.

\subsection{Scale separation and mixed neural representation}
% \subsection{Scaling mixture and separation of neural representations} \label{sec:scale-deformation}
Determining whether \textit{scale separation} exists in a complex dynamical systems is important factor that may be unknown \textit{a priori}. In place of scale separation, e.g. in self-organized criticality, there may be \textit{scale mixing}. If the function is locally-stable, then the system exhibits scale separation. However, if the transformation is unstable, the system exhibits scale mixing.

In multiscale renormalization group flows, one considers multiscale hierarchy of coarse-grained domains $\Omega_1,...,\Omega_S$. Coarse-graining commonly requires a metric in the domain to aggregate neighboring poinnts. A coarse-graining transformation is $f: \mathcal{X}(\Omega) \rightarrow \mathcal{Y}$ is locally stable if it can be factored as $f \approx f_s \circ C_s$, where $C_s : \mathcal{X}(\Omega) \rightarrow \mathcal{X}_s(\Omega_s)$ is a nonlinear coarse-graining transformation and $f_s: \mathcal{X}_s\left(\Omega_s\right) \rightarrow \mathcal{Y}$. While the transformation $f$ may depend on long-range interactions, locally-stable functions separate the ineractions across scales by propagating fine-grained interactions to coarse-grained scales.

\subsection{Multiscale representation stability}
% Fourier transform, Schur transform, Laplacian transform
Hierarchical multiscale representations facilitate the emergence of approximate symmetries, however, this requires a careful selection of multiscale transformations. A salient example is the instability the Fourier transform: a momentum-space representations given by the Fourier transform are unstable to high-frequency noise, while hybrid-space representations, such as wavelet transforms, are stable to deformations. 

Mallut (2012) argued that the Fourier transform is unstable under high-frequency deformations. Given an approximate translation $h(u) = u - \tilde{h}(u)$ with $\|\nabla \tau\|_{\infty}=\sup _{u \in \Omega}\|\nabla \tilde{\tau}(u)\| \leq \epsilon$,
$$
\frac{\|f(\rho(h) x)-f(x)\|}{\|x\|}=\mathcal{O}(1).
$$
where $f(x)=|\hat{x}|$ is the modulus of Fourier components
$\hat{x}(\omega)=\int_{-\infty}^{+\infty} x(u) e^{-i \omega u} \mathrm{~d} u$. Because the right-hand side is independent of $\epsilon$, the perturbation from a shift, the Fourier transform is unstable under deformations.

In contrast, localized filters that extract information from signals across multiple scales produces a family of locally stable features. A multiscale wavelet decomposition $W_\psi x$ is approximately equivariant to deformations,
$$
\frac{\left\|\rho(h)\left(W_\psi x\right)-W_\psi(\rho(h) x)\right\|}{\|x\|}=\mathcal{O}(\epsilon) .
$$
where the continuous wavelet transform
$\left(W_\psi x\right)(u, \omega)=\omega^{-1 / 2} \int_{-\infty}^{+\infty} \psi\left(\frac{v-u}{\omega}\right) x(v) \mathrm{d} v$ depends on a mother wavelet $\psi$. While multiscale wavelet representations approximately equivariant to deformations, discovering approximately invariant properties requires an integration over faster scales to slower scales.

The instability of the the Fourier transform to deformation extends to other momentum-space representations, such as the graph Fourier transform to spectral decompositions of the Laplacian matrix for complex networks and the Schur transform to the total angular momentum basis of spin systems.

% Occam's razor

% Variational representation learning

% Locally Stable Transformation
% Mesoscopic descriptions of neural systems may be constructed from the microscopic, complex network description. Coarse-graining transformations are key to constructing higher-order representations with locally stable scale separation. Deformation stability plays an important role in the selection of transformations---e.g., locally stable, wavelets, and Fourier. 

% \subsection{Scale separation in space and time}
% \subsection{Separation of Time Scales}

% Often, the exact symmetry-constrained transformation must be learned from data.

% Wavelets, local transformations

\section{Contributions}
In this chapter, I  developed foundational theoretical insights at the intersection of stochastic thermodynamics and representational geometry to address the challenge of modeling non-equilibrium neural systems. The primary contributions of this chapter include:

\begin{enumerate}
    \item \textbf{Unified Theoretical Framework}: Established a combined theoretical perspective for characterizing the representational geometry of neural systems. This unified framework characterizes symmetry-breaking phenomena and emergent scaling structures in non-equilibrium neural systems.
    \item \textbf{Symmetry and Equivariance in Neural Networks}: Offered approaches to constructing and relaxing symmetry constraints, focusing on permutation symmetries of neural networks. By exploring symmetries, approaches to modeling symmetry-breaking transformations in neural systems are identified.
\end{enumerate}

Overall, these contributions establish theoretical foundations in support of computational techniques presented in subsequent chapters, laying the groundwork for modeling multiscale neural dynamics.

\newpage
\section*{Appendix: Formalism for Exact Symmetry via Group Theory}
A \textit{group} is defined as a non-empty set $G$ with a binary composition operation and four properties---associativity, identity, inverse, and closure---which constrains the possible symmetries. 
A \textit{group action} of $G$ on $\Omega$ is defined as a mapping that associates a group element $g \in G$ and a point $u \in \Omega$ with some other point on the domain. %: $(g, u) \mapsto g . u$. 
Let $L_d$ denote the set of $d \times d$ general linear matrices. A \textit{representation} of the group is a linear group action or map $\rho: G \rightarrow \mathbb{R}^{d \times d}$ that assigns an invertible matrix $\rho(g) \in L_d$ to each element $g$ in the group.
%, where the dimension $N$ may be the dimensionality of the feature space, e.g. $N$ discrete time points or network nodes.
% Thus, a representation of a group $G$ is a map $\rho: G \rightarrow \mathbb{R}^{N \times N}$ assigning to each element $g \in G$ a matrix $\rho(g)$. 
If the map is many to one, it is known as a \textit{homomorphism}. If it is one-to-one, then the map is an \textit{isomorphism}. 

% Written as a representations, the group action on signals is defined as $\rho(g) x(u)=x\left(g^{-1} u\right)$

% If we have a group action on the domain, then there is a corresponding group action on the signal space: $(g.x)(u)=x\left(g^{-1} u\right)$. 

\subsubsection{Invariance and equivariance}
A transformation $f: \mathcal{X}(\Omega) \rightarrow \mathcal{Y}$ is \textit{invariant} with respect to a group $G$ if
$$f(\rho(g) x)=f(x)$$ for all $g \in G$ and $x \in \mathcal{X}(\Omega)$. A transformation $f: \mathcal{X}(\Omega) \rightarrow \mathcal{X}(\Omega)$ is \textit{equivariant} if
$$f(\rho(g) x)=\rho(g) f(x)$$ 
for all $g \in G$ and $x \in \mathcal{X}(\Omega)$.

\subsubsection{Equivalence and reducibility}
Two matrix groups are \textit{equivalent} if the groups are isomorphic and the corresponding elements under the isomorphism have the same character: the character of a matrix group $G \subset L_d$ is a function on the group defined by $\chi(g)=\operatorname{tr}(g)$ for $g \in G$. A matrix group $G \subset L_d$ is completely \textit{reducible} if it is equivalent to another matrix group $H$ in block diagonal form: all elements $m \in H$ have the form $\operatorname{diag}\left(m_1, m_2\right)$, for $m_1 \in L_{k_1}$ and $m_2 \in L_{k_2}$. If there is no such equivalence, the matrix group is \textit{irreducible}. Shur's lemma is a useful property of irreducible matrix groups.

\textit{Lemma 2.1}: (\textbf{Schur's lemma}) Let $G \subset L_d$ and $H \subset L_k$ be two matrix groups of the same order, $|G|=|H|$. If there exists a $k$ by $n$ matrix $S$ such that $S g_i=h_i S$ for some ordering of all elements $g_i \in G$ and $h_i \in H$, then either $S$ is the zero matrix, or $n=k$ and $S$ is a square nonsingular matrix. The following theorem connects irreducibility with characters

\textit{Theorem 2.1}: A matrix group $G$ is irreducible if and only if
$$
\frac{1}{|G|} \sum_{g \in G}|\chi(g)|^2=1
$$

\subsection*{Approximate symmetries and deformation stability}
% \subsection*{Deformation stability and broken symmetry complexity}
Global symmetries may be broken across multiple scales or critical transitions. However, because collective properties may be invariant or equivariant to local deformations and perturbative noise, a more relaxed notion of symmetry is required. Relaxed symmetries can be defined as the stability to deformations in terms of the distance between domains underlying signals or the complexity of perturbations away from a symmetry group $G$. 

Often, the domain itself is subject to deformations, as in the case in temporal graphs and dynamic manifolds. Let $d(\Omega,\tilde{\Omega})$ measure the distance between two instances $\Omega$ and $\tilde{\Omega}$ of the domain. When the two domains are equivalent, $d(\Omega,\tilde{\Omega})=0$. For example, $d(\Omega,\tilde{\Omega})$ may be the graph edit distance, measuring the minimal sequence of graph edits, or the Gromov-Hausdorff distance, measuring the minimal metric distortion between manifolds. A transformation $f$ is stable to \textit{domain deformations} or \textit{approximately invariant} if 
$$
\|f(x, \Omega)-f(\tilde{x}, \tilde{\Omega})\| \leq C\|x\| d(\Omega, \tilde{\Omega})
$$
where $C$ is a scalar.

Alternatively, the domain may be fixed and the signal is subject to deformations. Let $c(h)$ measure the complexity of the broken symmetry, such that $c(h)=0$ when $h \in G$. For example, the Dirichlet energy $c^2(h):= \int_{\Omega}\|\nabla h(t)\|^2  dt$ measures the elastic distortion by scalar field or, equivalently, the distance between $h$ and the translation group. 
% ---\cite{Wang2024} offers an attractive approach to discovering broken translational symmetry with relaxed group convolutions. 
A transformation $f$ is stable to \textit{signal deformations} or \textit{approximately invariant} if 
$$
\|f(\rho(h) x)-f(x)\| \leq C c(h)\|x\|, \forall x \in \mathcal{X}(\Omega).
$$
Similarly, a transformation $f$ is \textit{approximately equivariant} if 
$$
\|f(\rho(h) x)-\rho(h) f(x)\| \leq C c(h)\|x\|, \forall x \in \mathcal{X}(\Omega).
$$

\subsection*{Irreducible representations via the Schur transform}
The Schur basis for a $n$-particle and $d$-dimensional system is a generalization of the total angular momentum basis that is useful for exploiting the Schur-Weyl duality of symmetry under permutations $\mathcal{S}_n$ and collective linear operators~\citep{Bacon2006}. Instead of scaling exponential with the number of particles in the system, the number of degrees of freedom in the permutation invariant part of a density matrix scales polynomially with the number of particles.
The measurement is poly $\log(d)$. 

Consider a neural system of $n$ neurons, where each neuron is equipped with a local $d$-dimensional basis $[i], i=1 \ldots d$. The symmetric group $\mathcal{S}_n$ contains elements that give all possible orderings of the set of site indices $\{1, \ldots, n\}$. There are exactly $n!$ permutations, so $\mathcal{S}_n$ is a very large group.
$$
\mathbf{P}(\pi)[i_1 i_2 \cdots i_n]=[i_{\pi^{-1}(1)} i_{\pi^{-1}(2)} \cdots i_{\pi^{-1}(n)}]
$$
where $\pi \in \mathcal{S}_n$ and $[i_1 i_2 \ldots i_n ]$ is a shorthand for $[i_1] \otimes[i_2] \otimes \ldots [i_n]$. The matrix group is represented in the system by
$$
\mathbf{Q}(L)[i_1 i_2 \cdots i_n]=L[i_1] \otimes L[i_2] \otimes \cdots \otimes L[i_n]
$$
where $L \in \mathcal{L}_d$. Schur–Weyl duality relates irreducible finite-dimensional representations of the the symmetric group $\mathcal{S}_n$ and collective linear operations $\mathcal{L}_d$. These two actions commute, and in its concrete form, the Schur–Weyl duality asserts that under the joint action of the groups $\mathcal{S}_n$ and $\mathcal{L}_d$, the tensor space decomposes into a direct sum of tensor products of irreducible modules (for these two groups) that actually determine each other,
$$
\left(\mathbb{C}^d\right)^{\otimes n} \cong \bigoplus_{\lambda \in \operatorname{Part}[n, d]} \mathcal{Q}_\lambda \otimes \mathcal{P}_\lambda,
$$
The summands are indexed by the Young diagrams $D$ with $d$ boxes and at most $n$ rows, and representations $\pi_d^D$ of $\mathcal{S}_n$ with different $D$ are mutually non-isomorphic, and the same is true for representations $\rho_n^D$ of $\mathcal{L}_d$.
\chapter{Statistical Mechanics of Non-Equilibrium Neural Circuits}\label{chapter:geometric-thermodynamics}

Open neural systems exchange matter, energy, and information with external environments to support adaptive, nonequilibrium processes targeted at learning patterns in the world. The nonequilibrium thermodynamics of open neural systems may be understood by constructing a diffusion model of dissipative neural circuits---demonstrating that nonequilibrium steady-states imply asymmetric coupling neural networks---and coarse-graining heterogeneous neural populations to identify their effective dynamics---the nonlinearity in the spike-response of biological neurons may be captured by universal noise, likelihood models in the form of generalized linear models %(\ref{sec:glm}) 
% or geometric tensor products of the collective dynamics (master equation) of the binary spiking part of the system. 

% \begin{figure}[t]
% 	\includegraphics[width=1\textwidth]{figures/fig1.pdf}
% 	\caption{\small Representation of partially-observed neural circuits as a dissipative thermodynamic systems. (Left) A schematic representation of a neural circuit with sparse asymmetric connectivity. (Right) Dissipation-driven system implies fluctuations.}
% %	\caption{the refraction of light}
% \end{figure}\label{fig:geometry}

% Mathematical models of diffusion processes offer general analytical and computational frameworks for describing the statistical mechanics of partially-observed neural systems subject to different classes of spatiotemporal subsampling (Fig. 1).
% \section{Generalized Diffusion Model of Neural Networks}\label{sec:1-diffusion}
% \section{Diffusion dynamics in neural systems}
Analyzing the Fokker-Planck equation offers insight into the continuous Langevin diffusion dynamics of stochastic systems and neural networks. Diffusion is central to understanding the stochastic thermodynamics in fluctuation-driven neural systems: generative models~\citep{Rissanen2023,Dockhorn2022,Song2021}, stochastic gradient descent~\citep{Adhikari2023}, and non-equilibrium neural networks~\citep{Yan2013}. Dissipation-driven neural systems are governed by non-conservative forces that dissipate energy into the external environment across multiple temporal and spatial scales. 

Dissipative forces give rise to equal and opposite reaction forces that are manifested as fluctuations in open neural systems. These fluctuations exhibit non-trivial coupling that is independent of the coupling of neural circuits, but is instead dependent on external coupling to stochastic environmental variables: this external coupling is described by the diffusion tensor. Diffusion offers analytical and computational tools for describing the statistical mechanics of open neural systems. One of the major challenges in describing the statistical mechanics of neural systems is the presence of non-equilibrium steady-state probability flows originating from asymmetric network connectivity. Diffusion provides an analytical framework for the mathematical and numerical analysis of fluctuations in non-equilibrium neural circuits.

\begin{figure}[t]
	\includegraphics[width=1\textwidth]{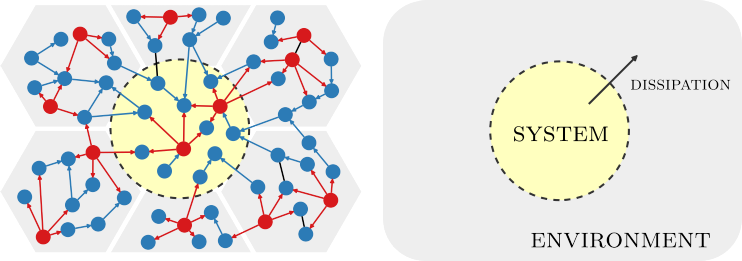}
	\caption{\small Relation of subsampled neural systems to the dissipative, non-equilibrium thermodynamics of open neural systems. (Left) Schematic representation of a neural circuit with sparse, signed asymmetric coupling. (Right) Dissipation from the neural system to the external environment drives multiscale fluctuations.}
\end{figure}\label{fig:geometry}

% subject to different subsampling schemes in space (Fig. \ref{fig:spatial-subsampling}). 

% \subsection{Deriving steady-state solutions of the Fokker-Planck equation}

% \subsection{Fokker-Planck formulation and non-equilibrium steady-states}
% \subsection{Fluctuation-Dissipation Relation in Spin-Glass Neural Networks}

% The theory of general diffusion processes offers a mathematical framework...
% \subsection{Asymmetric non-equilibrium neural circuits are dissipative}
% \subsection{Asymmetric neural networks are out-of-equilibrium}
% \subsection{Non-equilibrium Flows Emerge in Asymmetric Neural Networks}
% \subsection{Nonequilibrium steady-states imply asymmetric neural network coupling}\label{sec:ness}
% \subsection{Asymmetric Neural Networks are Out-of-Equilibrium}
% \subsection{Nonequilibrium Steady-State of Asymmetric Neural Networks}\label{sec:ness}
% \subsection{Fokker-Planck formulation and nonequilibrium steady-states}
% \subsection{Nonequilibrium steady-state $\implies$ asymmetric coupling}

Dissipative neural systems exchange energy with the environment and dissipate heat. Therefore, non-equilibrium steady-states will naturally result from dissipation-driven dynamics that decompose into potential-derived drift forces and fluctuation-driven diffusion modes. In non-equilibrium steady-states, persistent probability currents will manifest as fluctuation diffusion modes across space and time. Our approach is aimed at modeling these two phenomena. Langevin diffusion dynamics are described by a stochastic differential equation. Analysis of the corresponding Fokker-Planck equation has several implications: (1) asymmetric coupling gives rise to non-equilibrium steady-states and (2) free energy is a Lyapunov function.
\section{Equilibrium spin-glass neural networks in discrete-time}
% \subsection{Markov chains}
% \subsection{Discrete Glauber dynamics of stochastic bits}

Equilibrium spin-glass neural networks with Glauber dynamics have well-defined transition rates and satisfy detailed balance. A direct approach to modeling the circuit would take the activity of the population at a single moment in time as the configuration state of the circuit. Let $\sigma_i^\alpha(t) = {0,1}$ denote the activity of neuron i in population $\alpha$,  $w(\sigma_i^\alpha)$ be the transition rate (probability per unit time) that the neuron changes its state, and $P(\vec{\sigma},t)$ be the probability distribution of the state $\vec{\sigma} = { \sigma_i} (i = 1,...,N_{tot})$ at time $t$, where $N_{tot}$ is the total circuit size. 

\subsection{Master equation}
The master equation is
\begin{equation}
\frac{d}{dt} P(\vec{\sigma},t) - P(\vec{\sigma},t) \sum_i^{N_{tot}} w(\sigma_i) + \sum_i^{N_{tot}} P(\vec{\sigma}',t) w(1 - \sigma_i)
\end{equation}
where $\vec{\sigma}' = {\sigma_1,...,1-\sigma_i,...,\sigma_{N_{tot}}}$. The transition probabilities are:
\begin{equation}
w(\sigma_i ^\alpha) = \frac{1}{\tau_\alpha}[\sigma_i^\alpha - \Theta (h_i^\alpha)]^2
\end{equation}
\begin{equation}
w(\sigma_i ^X) = \frac{1}{2\tau_X}[1-(2\sigma_i^X-1)(2 m_X -1)]^2
\end{equation}
where $\Theta$ is the Heaviside (unit step) function, $m_X$ is the mean activity of neuron i from the external circuit (X), and $h_i^\alpha$ is the net afferent input to the neuron:
\begin{equation}
h_i^\alpha = \sum_\beta ^{X,E,I} \sum_j ^N J_{i j} ^{\alpha \beta} \sigma_j ^\beta - \theta_i ^\alpha.
\end{equation}
% The equilibrium solution is given by:
% \begin{equation}
% m_i^\alpha = \langle \Theta (h_i ^\alpha) \rangle
% \end{equation}
% \begin{equation}
% 2 r_{i j}^{\alpha \beta} = \langle \delta \sigma_i ^\alpha \delta \Theta(h_i ^\beta) \rangle +  \langle \delta \Theta(h_i ^\alpha) \delta \sigma_i ^\beta \rangle
% \end{equation}
% \begin{equation}
% 2 r_{i j}^{\alpha X} =  \langle \delta \Theta(h_i ^\alpha) \delta \sigma_j ^X \rangle
% \end{equation}
% where $r_{i j} ^{\alpha \beta} = \langle \delta \sigma_i ^\alpha (t) \delta \sigma _j ^\beta (t) \rangle$ for $\alpha_i \neq \beta_j$ and $\delta x = x - \langle x \rangle.$

% \subsection{Reservoir-coupled binary neural circuits}\label{sec:binary-networks}
\subsection{Entropy rates}\label{sec:binary-networks}
Consider the discrete-time state-space dynamics of a spin-glass neural network. With the transition matrix, one may estimate the total entropy rate of the stochastic system. The total entropy rate may be decomposed into an entropy production $\dot{S}^{i}(t)$ term intrinsic to the system and an entropy flow $\dot{S}^{e}(t)$ term describing the exchange of entropy with the environment: $\dot{S}(t) = \dot{S}^{i}(t) + \dot{S}^{e}(t)$. This distinction was first made by Ilya Prigogine in 1961~\citep{Prigogine1961}. In terms of the transition matrix, the two terms of interest are given by
\begin{equation}
\dot{S}^{i}(t) = \frac{k_B}{2}\sum_{ij,i'j'}J_{ij,i'j'}(t)\ln\frac{W_{ij,i'j'}p_{i'j'}(t)}{W_{i'j',ij}p_{ij}(t)}
\end{equation}
\begin{equation}
\dot{S}^{e}(t) = -\frac{k_B}{2}\sum_{ij,i'j'}J_{ij,i'j'}(t)\ln\frac{W_{ij,i'j'}}{W_{i'j',ij}}
\end{equation}
where $k_B$ is the Boltzmann constant and $J_{ij,i'j'}$ is the current from state $i'j'$ to $ij$. 
% Thus, we arrive at an estimate of $\dot{S}^{i}(t) = 3.554$ per second for the entropy production of the system; it is at equilibrium, so the total entropy production is zero and the entropy flow is equal and opposite to the entropy production. This approach has yielded an estimate for several entropy rates for the entire circuit, albeit, a simple example; importantly, this approach can be generalized to non-equilibrium neural circuits where the total entropy rate might not be balanced. 
Modeling state-space dynamics is obstructed by the curse of dimensionality. Several promising approaches to state-space dynamical modeling include Koopman operators from systems and control engineering, and tensor network from quantum many-body physics, and symmetry-preserving neural networks from geometric deep learning.

\section{Non-equilibrium Langevin dynamics in continuous-time}
The time-evolution of a stochastic neural network may be described by a stochastic differential equation,
\begin{equation} \label{1-eqn:ito-sde}
    d\mathbf{x} =\mathbf{f}(\mathbf{x},t)dt+\mathbf{G}(\mathbf{x},t) d\mathbf{w},
\end{equation}
where the state $\mathbf{x} \in \mathbb{R}^N$ represents activations and/or weights in neural networks, $\mathbf{f}(\cdot, t): \mathbb{R}^N \rightarrow \mathbb{R}^N$ is a drift term, $\mathbf{G}(\cdot, t): \mathbb{R}^N \rightarrow \mathbb{R}^{N \times M}$ is a diffusion term, and $\mathbf{w}$ is a $M$-dimensional standard Wiener process. Langevin diffusion dynamics describe a range of useful neural network models in which deterministic forces described by the drift coefficients $\mathbf{f}(\mathbf{x},t)$ add together with stochastic noise described by the diffusion coefficients $\mathbf{G}(\mathbf{x},t)$. 
% [TODO: Check that FR, I\&F models are allowed]

Diffusion allows for noisy input currents arising from subsampled observations with different statistical properties, e.g. Gaussian versus non-Gaussian. For instance, random subsampling of sparse neural networks gives rise to state-independent, homogeneous Gaussian noise $\mathbf{G}(t)$ whereas biased subsampling of the same networks gives rise to state-dependent, inhomogeneous non-Gaussian $\mathbf{G}(\mathbf{x},t)$ noise. 

Regarding the deterministic (internal) behavior of the subsampled populations, the forces $\mathbf{f}(\mathbf{x},t)$ can be realized by computational neuroscience models---generalized linear models, integrate-and-fire networks, firing rate models---and thermodynamic AI models---spin-glass/Monte Carlo sampling models, Bayesian neural networks, and generative diffusion models. Although the short-term behavior and transient dynamics are noteworthy, the primary focus is typically on the steady-state forces, probability distributions, and probability currents, which can be examined using the Fokker-Planck equation.

% \section{Nonequilibrium Flows of Asymmetric Neural Networks}\label{sec:1.2}
% \subsection{Fokker-Planck equation}\label{sec:ness}% formulation: steady-state distributions}

% \subsection{Fokker-Planck}
% Brunel, Sparse

\subsection{Fokker-Planck equation}
% \subsubsection{Fokker-Planck equation: homogeneous vs. inhomogeneous noise}
The Fokker-Planck equation describes the time evolution of the probability density $p(\mathbf{x}_t,t)$, according to a continuity equation that is based on the conservation law of probability

\begin{equation} \label{1-eq:continuity-eqn}
   \frac{\partial p(\mathbf{x},t)}{\partial t}=-\nabla 
   \cdot \mathbf{J}(\mathbf{x}, t),
\end{equation}
where $\mathbf{J}(\mathbf{x}, t)=\mathbf{f}(\mathbf{x},t) p(\mathbf{x},t)- \nabla \cdot \left[\mathbf{D}(\mathbf{x}, t) p(\mathbf{x},t) \right] $ is a vector of probability currents out of state $\mathbf{x}$; $\mathbf{D}(\mathbf{x}_t,t)=\frac{1}{2} \mathbf{G}(\mathbf{x}_t, t) \mathbf{G}(\mathbf{x}_t, t)^{\boldsymbol{\top}}$ is the diffusion matrix.
\subsubsection{Stationary state solution}
% \subsubsection{Steady-states: equilibrium vs. nonequilibrium}

%[TODO: Contrapositive]

% By the conservation law of probability, we have a continuity equation
% \begin{equation} \label{eq:continuity}
%    \frac{\partial p(\mathbf{x},t)}{\partial t}+\nabla \mathbf{J}(\mathbf{x}, t)=0, 
% \end{equation}
% where $\nabla \mathbf{J}(\mathbf{x}, t)$ is the currents of the probability current out of state $\mathbf{x}$. Comparing Equations \ref{eq:continuity} and \ref{eq:continuity}, we deduce
% \begin{equation} \label{eq:3.3}
%     \mathbf{J}(\mathbf{x}, t) = \mathbf{f}(\mathbf{x},t)p(\mathbf{x},t)-\nabla(\mathbf{D}(\mathbf{x}_t,t)p(\mathbf{x},t))\\
% \end{equation}

% A steady-state probability distribution, $p^s(\mathbf{x})$, implies $\partial p^s(\mathbf{x},t)/\partial t=0$ and $\nabla \mathbf{J}^s(\mathbf{x}, t)=0$, i.e. steady-state probability currents $J^s(\mathbf{x})$ are \textit{divergence-free}. There are two possible scenarios: if $\mathbf{J}^s(\mathbf{x},t)=0$, then the system is maintained at equilibrium by forces acting along the gradient of an potential and $p^s(\mathbf{x})$ corresponds to the Boltzmann distribution; if $\mathbf{J}^s(\mathbf{x},t)\neq 0$,  then the system is maintained at a nonequilibrum steady state according to an potential-like factor and the steady-state probability currents $\mathbf{J}^s(\mathbf{x})$ and $p^s(\mathbf{x})$ is a nonequilibrum steady-state probability distribution.

For a stationary probability distribution, $p^s(\mathbf{x})$, in the long time limit---with $\mathbf{D}(\mathbf{x},t) \rightarrow \mathbf{D}(\mathbf{x})$ and $\mathbf{J}(\mathbf{x},t) \rightarrow \mathbf{J}^s(\mathbf{x})$---the effective force on a state $\mathbf{x}$ is
% steady-state probability flows are divergence-free $\nabla \cdot \mathbf{J}^s(\mathbf{x})=0$.
% There are two possible scenarios: if $\mathbf{J}^s(\mathbf{x},t)=0$, then the system is maintained at equilibrium by forces acting along the gradient of a potential and $p^s(\mathbf{x})$ corresponds to the Boltzmann distribution; if $\mathbf{J}^s(\mathbf{x},t)\neq 0$, then the system is maintained at a nonequilibrum steady state according to an potential-like term and the steady-state probability currents $\mathbf{J}^s(\mathbf{x})$ and $p^s(\mathbf{x})$ is a nonequilibrum steady-state probability distribution. 
% To simplify the theoretical analysis, consider state-independent homogeneous noise $g(t)$, $\mathbf{D}(t)=g^2(t)/2$.
% If $\mathbf{J}(\mathbf{x},t)\neq0$, t
% The system converges to a nonequilibrium steady-state probability distribution $p^s(\mathbf{x})$
\begin{align*}
    \mathbf{f}(\mathbf{x})
    &=\nabla \cdot \left[\mathbf{D}(\mathbf{x})  p^s(\mathbf{x})\right] / p^s(\mathbf{x})+\mathbf{J}^s(\mathbf{x}) / p^s(\mathbf{x})\\
    % &=\nabla \cdot \mathbf{D}(\mathbf{x}) + \mathbf{D}(\mathbf{x}) \nabla p^s(\mathbf{x}) / p^s(\mathbf{x}) + \mathbf{J}^s(\mathbf{x}) / p^s(\mathbf{x})\\
    &=\nabla \cdot \mathbf{D}(\mathbf{x}) - \mathbf{D}(\mathbf{x}) \nabla U(\mathbf{x}) + \mathbf{v}^s(\mathbf{x}).
    % -\mathbf{D}(\mathbf{x},t) \nabla\left(-\ln p^s(\mathbf{x})\right)+\mathbf{J}^s(\mathbf{x}) / p^s(\mathbf{x})\\
    % &=-\mathbf{D}(\mathbf{x},t) \nabla\left(-\ln p^s(\mathbf{x})\right)+\mathbf{J}^s(\mathbf{x}) / p^s(\mathbf{x})\\
    % &=-\boldsymbol{\mu}(\mathbf{x},t) \nabla\left(-D\ln p^s(\mathbf{x})\right)+\mathbf{J}^s(\mathbf{x}) / p^s(\mathbf{x})\\
    % &=-\boldsymbol{\mu}(\mathbf{x},t) \nabla_{\mathbf{x}} E(\mathbf{x})+\mathbf{J}^s(\mathbf{x}) / p^s(\mathbf{x}),
\end{align*}
where $U(\mathbf{x})$ is a pseudopotential satisfying $\nabla U(\mathbf{x}) = \nabla p^s(\mathbf{x}) / p^s(\mathbf{x})$
and $\mathbf{v}^s(\mathbf{x}) = \mathbf{J}^s(\mathbf{x}) / p^s(\mathbf{x})$ is a velocity field. Equilibrium states satisfy detailed balance, $\mathbf{J}^s(\mathbf{x})=0$, while non-equilibrium states have persistent probability currents $\mathbf{J}^s(\mathbf{x}) \neq 0$ and velocity field $\mathbf{v}^s(\mathbf{x}) \neq 0$.

In generalized Boltzmann equilibrium, two conditions are satisfied:
\begin{itemize}
    % \item Divergence condition: there is a scalar function, $A(\mathbf{x})$, such that $\nabla \cdot \left[A(\mathbf{x}) \mathbf{D}(\mathbf{x},t)\right] = 0$.
    \item Divergence condition: $\nabla \cdot \mathbf{D}(\mathbf{x}) = 0$.
    \item  Einstein relation: $\boldsymbol{\mu}(\mathbf{x}) = \beta \mathbf{D}(\mathbf{x})$ with proportionality constant $\beta$.
\end{itemize}
Given these conditions, detailed balance $\mathbf{J}^s(\mathbf{x})=0$ is satisfied by $U(\mathbf{x}) = \beta E(\mathbf{x})$,
% $U(\mathbf{x}) = \beta E(\mathbf{x}) - \log A(\mathbf{x})$, 
up to a constant, giving Boltzmann distributed probabilities: $p^s(\mathbf{x}) = \exp[-\beta E(\mathbf{x})]/Z$, where $Z = \int d\mathbf{x} p^s(\mathbf{x})$ is the partition function and $\beta = 1/k_B T$ is the inverse temperature.

Non-equilibrium steady-states have persistent probability currents $\mathbf{J}^s(\mathbf{x}) \neq 0$ and velocity field $\mathbf{v}^s(\mathbf{x}) \neq 0$., which means that steady-state forces cannot be written entirely in terms of a gradient of a potential energy function.

\subsection{Non-equilibrium implies asymmetric connectivity}\label{sec:1-ness}
% Non-equilibrium steady-state forces correspond to asymmetric coupling and entropy production
 % separate into symmetric and asymmetric coupling components
% For symmetrically coupled spin-glass neural networks, such as Hopfield neural networks and Boltzmann machines, the forces are given by
Consider a Hopfield neural circuit~\citep{Hopfield1984}, where each neuron's time-evolution is described by
\begin{align}
f_i=\dot{x}_i=\sum_j T_{ij} s(x_j)-x_i-h_i ,
\end{align}
where $x_i$ is the state of each neuron, $h_i$ is an external field, $T_{ij}$ represents the synaptic coupling from neuron $j$ to $i$, and $s_j(\cdot)$ is the sigmoid transfer function.

In a symmetrically-coupled circuit ($T_{ij}=T_{ji}$), the forces can be expressed as a gradient of an energy function~\citep{Yan2013} % Hopfiled
$$
\mathbf{f}(\mathbf{x})=-\mathbf{A}(\mathbf{x}) \nabla E(\mathbf{x}),
$$
where $A_{i j}=\delta_{i j}/{s'(x_i)}$ and
$E(\mathbf{x})=-\sum_i h_i s(x_i)- \frac{1}{2} \sum_{i \neq j } T_{ij} s(x_i) s(x_j) + \sum_i \int_0^{x_{i}}y s'(y)dy.$
% $$E(\mathbf{x})=-\\mathbf{h}^{\top}s(\mathbf{x})-s(\mathbf{x})^{\top} \mathbf{T}_{sym} s(\mathbf{x})/2,$$
% In general, symmetric coupling implies equilibrium forces, because an energy function may be defined. 
% This is because nonequilibrium forces in \textit{asymmetrically}-coupled neural networks cannot be expressed as a gradient of an energy function.
When these deterministic forces are subjected to stochastic noise described by diffusion coefficients $\mathbf{G}(\mathbf{x},t)$, the symmetrically-coupled circuits converge to equilibrium. Because the forces can be written entirely in terms of a gradient of a potential energy function, symmetric coupling is a sufficient condition for equilibrium.

% Assuming homogeneneous noise $G(t)$, the driving force decomposes as
% $$
% \mathbf{f}(\mathbf{x},t)=\mathbf{J}^s(\mathbf{x}) / p^s(\mathbf{x})-\mathbf{D}(t) \nabla E(\mathbf{x}).
% $$
% By comparing this equation to the force in the symmetric scenario, it becomes evident that the term $J^s(\mathbf{x}) / p^s(\mathbf{x})$ must arise from the asymmetric component of the force.
The contrapositive is that non-equilibrium forces imply asymmetric coupling. In the case of an asymmetrically-coupled circuit, non-equilibrium steady-state forces cannot be written entirely in terms of a gradient of a potential energy function; there is an additional term arising from the the probability flux. However, the symmetrically-coupled equilibrium case provides intuition for separating non-equilibrium steady-states forces into two components arising from symmetric and asymmetric coupling:
\begin{equation}
f_i=\underbrace{\sum_j T^s_{ij} s(x_j)-x_i-h_i}_{\text{\normalsize symmetric~$f^s_i$}}+\underbrace{\sum_jT^a_{ij} s_j(x_j)}_{\text{\normalsize asymmetric~$f^a_i$}}.
\end{equation}
with $\mathbf{T}^s = \mathbf{T}\mathbf{T}^{\top}$, corresponding to the co-citation network, and $\mathbf{T}^a = \mathbf{T} - \mathbf{T}^s = \mathbf{T} (\mathbf{1} - \mathbf{T})$. Subject to stochastic noise described by diffusion coefficients $\mathbf{G}(\mathbf{x},t)$, the neural circuit converges to a non-equilibrium steady-state with persistent probability currents $\mathbf{J}^s(\mathbf{x}) \neq 0$. Decomposition into symmetric and asymmetric parts recovers the energy function for the equilibrium case and isolates the non-equilibrium forces: $\mathbf{J}^s(\mathbf{x}) / p^s(\mathbf{x})  =  \mathbf{T}^a s(\mathbf{x})$.
% \[
% \underbrace{u'-P(x)u^2-Q(x)u-R(x)}_{\text{=0, since~$u$ is a particular solution.}}
% \]
% with the asymmetric component shifted to the right. We developed a symmetric connection matrix $ST$ structured as follows:
% $$
% ST_{ij} = a *\left(T_{i j}+T_{i i}+b *\left(T_{i j}+T_{i i}\right)^2\right)
% $$
% with constants $a$ and $b$. The matrix follows the property $S T_{i j}=S T_{j i}$. The **asymmetric** component is given by $\mu S T_{i j}=T_{i j}-S T_{i j}$. The resulting symmetric portion of the driving force
% $$
% f_i(symmetric)=\frac{1}{C_i}\left(\sum_{j=1}^N S T_{i, j} f_j\left(u_j\right)-\frac{u_i}{R_i}+I_i\right)
% $$
% exhibits the typical structure of a Hopfield model. 

\subsection{Asymmetric coupling yields nonequilibrium entropy production}
While the forces associated with the symmetric component of the network coupling correspond to gradient descent in an energy landscape, the forces associated with the asymmetric component of the network coupling are correspond to stochastic noise fluctuations with an entropy production.
We calculate the entropy production as a line integral~\citep{Adhikari2023,Seifert2012}.
\begin{align*}
    \dot{S}^{\text{tot}} 
    &=\int_\Omega p^s(x) \sum_{i=1}^d \mathbf{J}_i^s(\mathbf{x}) \frac{\partial}{\partial x_i}\left(\frac{\mathbf{J}_i^s(\mathbf{x})}{\mathbf{D}_{i i}(\mathbf{x}) p^s(\mathbf{x})}\right) d\mathbf{x}
    = \int_\Omega \frac{\mathbf{J}^s(\mathbf{x})^\top \mathbf{D}^{-1}(\mathbf{x}) \mathbf{J}^s(\mathbf{x})}{p^s(\mathbf{x})}d\mathbf{x}\\
\end{align*}
The line integral along $\Omega$ describes the entropy production rate in a probabilistic sense, integrating over all possible system states. This integral gives the total entropy production rate in terms of the steady-state probability current and the system's diffusion properties.

Defining the vector field $\mathbf{v}(\mathbf{x}(t)) \coloneq \mathbf{D}^{-1}(\mathbf{x}) \mathbf{J}^s(\mathbf{x}) / p^s(\mathbf{x})  = \mathbf{D}^{-1}(\mathbf{x}) \mathbf{T}^a \mathbf{s}(\mathbf{x})$, the path-integral over stochastic trajectories is 
\begin{align*}
    \dot{S}^{\text{tot}} = 
&= \int_{\Omega}\left[\mathbf{T}^a s(\mathbf{x})\right]^T \mathbf{D}^{-1}(\mathbf{x})\left[p^s(\mathbf{x}) \mathbf{T}^a s(\mathbf{x})\right] d \mathbf{x}=\int_{\Omega} p^s(\mathbf{x})\left[\mathbf{T}^a s(\mathbf{x})\right]^T \mathbf{D}^{-1}(\mathbf{x}) \mathbf{T}^a s(\mathbf{x}) d \mathbf{x}\\
&=\int_{t_0}^{t_f} \mathbf{v}(\mathbf{x}(t)) \cdot \dot{ \mathbf{x}}(t) d t = \int_{t_0}^{t_f} \mathbf{D}^{-1}(\mathbf{x}(t))  \mathbf{T}^a s(\mathbf{x}(t)) \cdot |\dot{ \mathbf{x}}(t)|  d t.
\end{align*}
where $\mathbf{x}(t_0)$ and $\mathbf{x}(t_f)$ give the endpoints of $\Omega$. The path-integral form of the entropy production is in terms of individual stochastic trajectories, using the relationship between probability currents and system dynamics, where $\{\mathbf{x}(t) \}$ is a collection of sample paths. This form expresses entropy production in terms of time evolution, linking the macroscopic thermodynamic forces to microscopic trajectory-based dynamics. This transformation is common in stochastic thermodynamics, where entropy production can be computed either from steady-state probability distributions or from observed system trajectories.

\section{Contributions}
\begin{itemize}
    \item \textbf{Analytical Characterization of Non-Equilibrium Systems}: Derived  analytical correspondences connecting symmetric and asymmetric coupling to equilibrium-like energy functions and non-equilibrium entropy production terms in neural systems. These correspondences clarify how asymmetry in neural connectivity shapes non-equilibrium steady-states.
\end{itemize}

% \part{Numerical}
% \part{Applications}
\chapter{Latent Graph Diffusion-Based Laplacian Renormalization}\label{chapter:lgd}

Spectral graph clustering techniques that rely on the eigenvalue decomposition of a graph's Laplacian matrix are interpreted as generating a momentum-space renormalization group flow in the space of metagraphs. In the context of analyzing large-scale neural recording data, this approach depends on a latent graph inferred by statistical models, such as generalized linear models. Numerical simulations of community-clustered networks are used as a proving ground for co-developing diffusion-based latent graph inference and Laplacian renormalization.

\section{Introduction}
Inferring the collective properties of complex neural systems is obstructed by the subsampling problem~\citep{Levina2022,Levina2017}. 
% Spatial subsampling leads to systematic biases in the discovery of coupling dynamics and latent graphs. 
One promising approach leverages renormalization group techniques from statistical physics to identify scale-invariant properties by iteratively coarse-graining subsampled neural networks~\citep{Meshulam2019}. While correlation-based renormalization effectively coarse-grains neural populations according to functional connectivity, other renormalization group techniques may be applied to the coarse-graining of neuronal networks. For example, the Laplacian renormalization scheme for complex networks relies on graph spectral methods to generate a renormalization group flow in the space of metagraphs~\citep{Villegas2023}.

Laplacian renormalization operates on the latent graph of complex networks, which must be inferred from observations if it is unknown. Generalized linear models are widely used to infer latent graphs and neuronal firing statistics observed in large-scale neural data~\citep{Tseng2022, Mahuas2020, Pillow2008, Paninski2004}. Although network reconstruction is NP-hard, recent work has been aimed at inferring network connectivity from event timing patterns~\citep{Casadiego2018} or mechanical models~\citep{Ladenbauer2019}. However, it may be more practical to extract community-level features and heterogeneous populations from neural data.

Generalized linear models not only infer network connectivity from observed neural activity, but also excitatory and inhibitory neuronal types.
However, subsampling effects may degrade model performance: suggesting the need to strengthen prior noise models. Diffusion has been shown to improve latent graph learning~\citep{Gasteiger2019}, which suggests alternative deep learning approaches using graph neural networks~\citep{Velickovic2018,Kipf2018}.
% One may also argue for more expressive graph neural network models~\citep{}, because diffusion improves latent graph learning. 
% The generalized diffusion framework includes a heat kernel, which is of interest for coarse-to-fine generative models with graphs~\citep{Rissanen2023}. 
Self-supervised graph learning is an attractive approach that accounts for the underlying permutation symmetry of graphs. When combined with the translational symmetry of convolutional filters in generalized linear models, one may construct a joint geometric deep learning model of large-scale neural recording data.

The key contributions are as follows:
\begin{itemize}
    \item Applied the diffusion-based renormalization approach to the latent graph learning problem in neural data.
    \item Demonstrated how diffusion-based renormalization infers latent graphs of community-clustered neural networks.
    \item Assessed how robust spectral characteristics of latent metagraphs are to latent graph inference.
\end{itemize}

\section{Methods}
\subsection{Leaky integrate-and-fire model}
Leaky integrate-and-fire models treat each neuron as an RC circuit operating in the overdamped regime: without external input, the membrane voltage decays exponentially to a resting potential $V_L$  according to the membrane time constant $\tau = R_m C_m$. The subthreshold dynamics of the membrane voltage $V$ are given by the ordinary differential equation
$$
\dot V(t)=-\frac{1}{\tau}\left [(V(t)-V_L)+R_mI(t) \right]
$$
where $I(t)$ is an externally injected current. If the voltage reaches a threshold voltage $V_{\text th}$ the neuron spikes and the voltage is reset to $V_R$ for an absolute refractory period $\tau_r$.

% \subsection{Statistical description of neuronal firing in generalized linear models}
Consider the stochastic dynamics of a leaky integrate-and-fire neuron with noisy current input $\eta(t)$ given by a stochastic differential equation
$$
dV(t)=-\frac{1}{\tau}\left [(V(t)-V_L)+R_mI(t) \right]dt + \eta(t) dw
$$
where $w$ is a one-dimensional standard Wiener process. Fluctuations lead to a diffusion in firing times, counts, and intervals. These firing statistics of a stochastic spiking neuron may be approached with generalized linear models.

Linear-nonlinear Poisson (LNP) cascades model neural spiking as a count process, where the probability of observing $k$ spikes in a discrete time interval $\Delta$ is given by
$$
P(k)=\frac{(\Delta \lambda)^k}{k!} e^{-\Delta \lambda}
$$
where the firing rate $\lambda$ is a nonlinear function of a linear filter $\mathbf{k}$ operating on an observed spike train $\mathbf{x}$  : $f(\mathbf{k} \cdot \mathbf{x})$. In practice, convolutional filters are constructed from a basis of raised cosine functions~\citep{Tseng2022,Pillow2008}.
% In the point process formulation, the probability that a neuron fires a spike depends on the instantaneous difference between the membrane voltage and threshold, $p_{\text {fire}} = 1 - \exp [V(t) -V_{\text th}]$.

% Generalized linear models 
% point and count process models describe the likelihood of 

% \subsection*{Escape rates, spike-response, generalized linear models, and linear-nonlinear Poisson}

% \subsubsection*{Balanced networks}

% \subsection*{Diffusion picture of generalized linear model}
% % \subsubsection{Integrate-and-fire as Langevin diffusion dynamics}

% $$
% \tau \dot{\mathbf{v}} = -\mathbf{v}+RI + \eta(t)
% $$
% % \subsection{Generalized Linear Models (GLM)}
% Convex optimization problem for exponential function.

% \subsubsection*{Spike-Response GLM}
% Escape rate as an exponential function:

% \subsubsection*{Linear-Nonlinear Poisson GLM}

% \textit*{Cosine Bump Basis Function}

% \section*{Appendix B: Numerical simulations}
\subsection{Clustered networks of leaky integrate-and-fire neurons}
\begin{figure*}
  \centering
  \includegraphics[width=\textwidth]{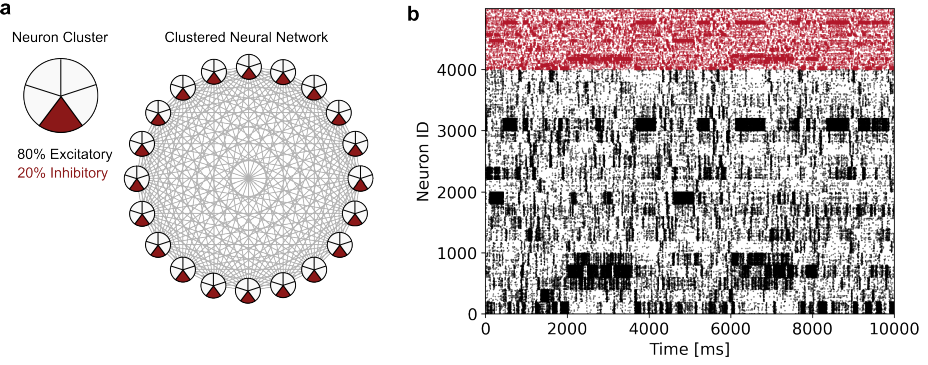}
  \caption{Numerical simulations of a biophysically-plausible clustered neural network. \textbf{a}, Each neuron cluster contains excitatory and inhibitory neurons in proportion to empirical observations of the cortical circuits. \textbf{b}, Raster plot of neuronal spike trains shows neural activity is asynchronous.}
  \label{fig:clustered-network}  
\end{figure*}

Clustered networks of leaky integrate-and-fire neurons were numerically simulated with balanced network parameters, as described previously~\citep{Rostami2024}. Each cluster of neurons contains excitatory and inhibitory neurons in proportion to empirical observations of the cortical circuits (Fig.~\ref{fig:clustered-network}a). Individual neuronal dynamics evolve according to the ordinary differential equation
$$
\dot V_i(t)=\frac{-\left[V_i(t)-V_L\right]}{\tau}+\frac{I(t)+I_{\mathrm{syn}}(t)}{C_m}
$$
where $I(t)$ is an externally injected current and $I_{\mathrm{syn}}(t)$ is the synaptic current input which evolves according to
$$
\tau_{\mathrm{syn}} \frac{d I_{\mathrm{syn}}^i}{d t}=-I_{\mathrm{syn}}^i+\sum_j J_{i j} \sum_k \delta\left(t-t_k^j\right)
$$
where $t_k^j$ is the arrival time of the $k$th spike from presynaptic neuron $j$ and $\delta$ is the Dirac delta distribution.

\section{Results}
We explore the diffusion-based Laplacian renormalization group procedure for community-clustered neural networks. Spectral graph analysis reveals higher-order graph features and network diffusion modes that correspond to network structures known \textit{a priori}. Robustness of spectral graph techniques to latent graph inference is studied by fitting a generalized linear model to numerical simulations of neural spiking activity. Latent graphs inferred by generalized linear models are shown to contain community features and excitatory - inhibitory interactions, and spectral graph analysis partially recovers dominant network diffusion modes for community-clustered neural networks.

\subsection{Spectral graph features of community-clustered neural networks}
\begin{figure*}[h]
  \centering
  \includegraphics[width=\textwidth]{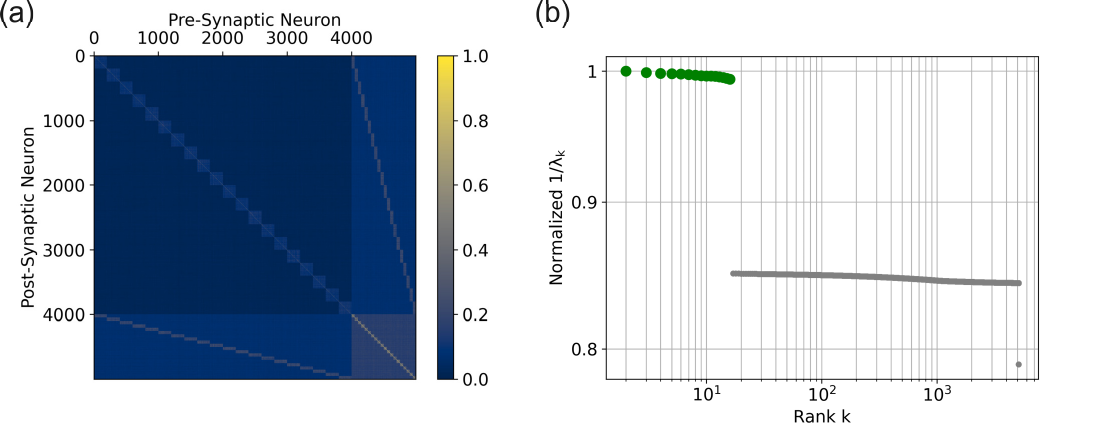}
  \caption{Spectral graph features of clustered neural network. (a), Soft-adjacency matrix $\mathbf{A}=\mathbf{C}\mathbf{C}^{\top}$ representing the symmetric component of the connectivity, i.e. the co-citation network. (b), Spectral decomposition of the Laplacian matrix $\mathbf{L}=\mathbf{D} - \mathbf{A}$ reveals an abrupt transition between the top-$k$ components for $k= n_\text{clusters}$.}
  \label{fig:spectral-graph-analysis}  
\end{figure*}

% \subsection{Latent graphs inferred from stochastic graph diffusion models}
% Cluster neuronal networks~\citep{Rostami2024}. Numerical simulations.
For a clustered neural network, spectral graph analysis is expected to identify structures at the coarser scales of populations and communities. Network connectivity is symmetrized by computing a soft-adjacency matrix $\mathbf{A}=\mathbf{C}\mathbf{C}^{\top}$ representing the symmetric component of the connectivity, i.e. the co-citation network (Fig.~\ref{fig:spectral-graph-analysis}a). Eigenvalue decomposition of the Laplacian matrix  $\mathbf{L}=\mathbf{D} - \mathbf{A}$, where $\mathbf{D}$ is the degree matrix, reveals an abrupt transition in the top-$k$ eigenvalues at $k= n_\text{clusters}$ (Fig.~\ref{fig:spectral-graph-analysis}b).

\begin{figure*}[h]
  \centering
  \includegraphics[width=\textwidth]{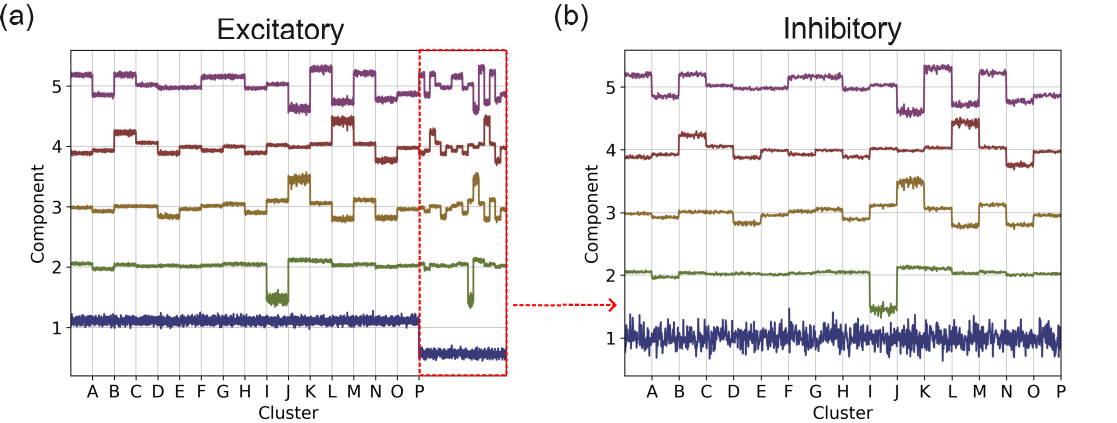}
  \caption{Clustering of the neural network in the top five eigenvectors of the Laplacian matrix. (a), Network modes divide the network into excitatory and inhibitory populations in the first component, and community-level features in the second through fifth components. (b), Community-level features are consistent across inhibitory and excitatory populations.}
  \label{fig:laplacian-components}  
\end{figure*}

Visualizing the eigenvectors of the Laplacian matrix leads to identifying higher-order features of the graph. Population and community-level clustering is apparent in the top five eigenvectors (Fig.~\ref{fig:laplacian-components}a). The first component corresponds to the second-smallest eigenvalue, known as the Fiedler value of the graph, and partitions the graph into two subgraphs along a cut that minimizes the weight of the cut: the Fiedler eigenvector cuts the clustered neural network into excitatory and inhibitory populations. Community-level structure is apparent in the next eigenvector and is consistent across excitatory and inhibitory populations (Fig.~\ref{fig:laplacian-components}b).

With respect to diffusion-based renormalization, the corresponding eigenvectors are associated with the dominant modes that govern information network diffusion. Constructing a lower-rank Laplacian matrix $L_k=\sum^k_i \lambda_i \mathbf{v}_i \mathbf{v}_i^\top$ for principal eigenvalues $\lambda_i$ and eigenvectors $\mathbf{v}_i$ is tantamount to Wilson's momentum-space renormalization group for complex networks. Laplacian renormalization group flows iteratively achieve coarse-grained metagraphs, which is desirable for understanding the hierarchical structure of the neocortical networks in the brain.

Inferring latent graphs from neural spiking activity is a technical capability that is being actively developed. In the next section, we explore how well a generalized linear model performs latent graph inference. We apply spectral graph analysis to the inferred latent graphs to assess the robustness of the Laplacian renormalization group approach.

\subsection{Robustness to latent graphs inference from neural activity}
\begin{figure}[h!]
  \centering
  \includegraphics[width=\textwidth]{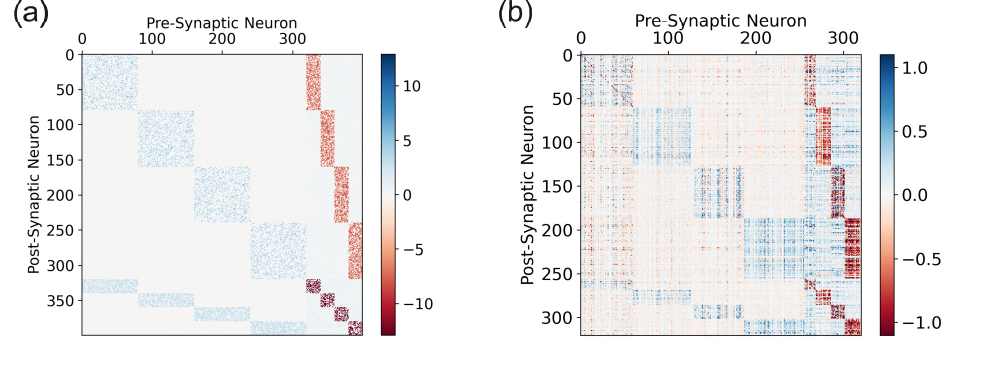}
  \caption{Latent graph inferred by a LNP generalized linear model contains key properties of the network architecture. (a) True network connectivity of a community-clustered neural network has synaptic types and community-level clustering. (b) Inferred network connectivity of a LNP generalized linear model has synaptic types, community-level clustering, and absolute refractoriness.}
  \label{fig:inferred-network}
\end{figure}

We assess the robustness of the Laplacian renormalization group approach to latent graph inference methods using generalized linear models.
Fitting linear-nonlinear Poisson (LNP) generalized linear models to neural spiking activity leads to the recovery several key network properties of a community-clustered neural network (Fig.~\ref{fig:inferred-network}): (1) synaptic types corresponding to excitatory and inhibitory interactions; (2) community-level clustering; (3) absolute refractoriness of individual neurons are captured on the diagonal of the reconstructed matrix. Up to a constant rescaling factor, the reconstructed connectivity appears to depict an accurate representation of the underlying connectivity structure.

\begin{figure}[h!]
  \centering
  \includegraphics[width=\textwidth]{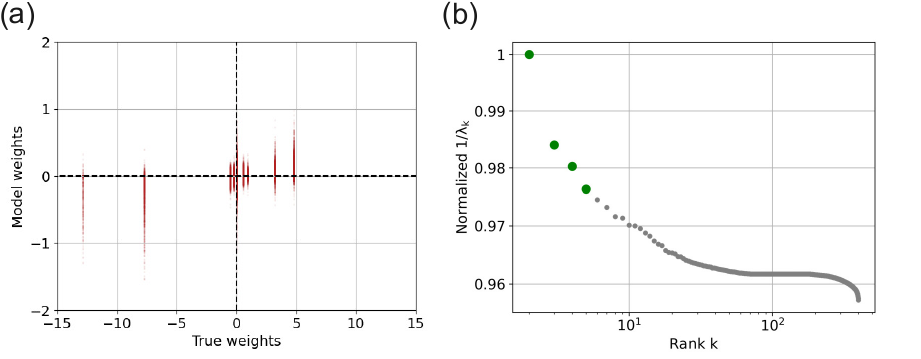}
  \caption{Qualitative properties of the network connectivity are degraded in the process of LNP-based latent graph inference. (a) Inferred model weights are plotted against true model weights. (b) Spectral components of the inferred graph Laplacian matrix.}
  \label{fig:reconstruction-quality}
\end{figure}

However, closer analysis of the qualitative properties of the network connectivity reveals important degradations in the process of LNP-based latent graph inference (Fig.~\ref{fig:reconstruction-quality}). Although LNP-based latent graph inference recovers most of the strongest excitatory and inhibitory synaptic weights, performance degrades as the synaptic weights become weaker (Fig.~\ref{fig:inferred-network}a). Performance on weakly-coupled weights near zero is particularly poor and spurious synaptic weights at zero have been falsely inferred. Moreover, there is no clear transition in the spectral components of the Laplacian matrix, making the definition and counting of clusters ambiguous (Fig.~\ref{fig:inferred-network}b).

\begin{figure}[h!]
  \centering
  \includegraphics[width=\textwidth]{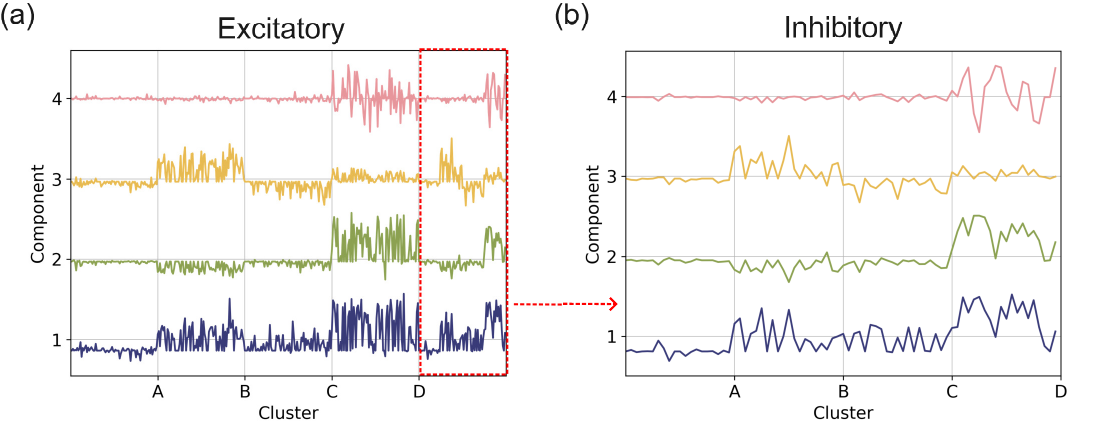}
  \caption{Eigenmodes of the network. (a) True network connectivity of a community-clustered neural network. (b) Inferred network connectivity of a linear-nonlinear Poisson generalized linear model.}
  \label{fig:laplacian-components-reconstructed}
\end{figure}

At the community-level, eigenmodes of the network contain variations with different community clusters (Fig.~\ref{fig:laplacian-components-reconstructed}a). Note, the Fiedler eigenvector does not split the network into excitatory and inhibitory populations, as observed previously; however, the network eigenmodes are consistent across excitatory and inhibitory populations (Fig.~\ref{fig:laplacian-components-reconstructed}b).

\section{Contributions}
In this chapter, I demonstrated latent graph diffusion (LGD) for inferring hidden structures from large-scale neural data.
This algorithm combines generalized linear modeling and diffusion-based spectral analysis to address challenges associated with subsampled neural systems. The primary contributions include:

\begin{itemize}
    \item \textbf{Latent Graph Diffusion Algorithm}: Developed and applied the latent graph diffusion algorithm for inferring and analyzing latent graph structures from neural data. This approach leverages spectral graph clustering techniques, interpreted as momentum-space renormalization flows, to systematically coarse-grain neural networks.
    \item \textbf{Community-Level Latent Graph Inference}: Demonstrated the efficacy of the diffusion-based renormalization approach for discovering community structures within clustered neural networks. Spectral decomposition of the inferred Laplacian matrices revealed higher-order community and population-level features consistent with known network architectures.
    \item \textbf{Robustness of Spectral Features}: Assessed the robustness of the latent graph diffusion algorithm to the inaccuracies inherent in generalized linear models. Although generalized linear models captured key network properties (excitatory-inhibitory interactions, community structures, and refractoriness), spectral analysis revealed limitations in accurately recovering weak synaptic interactions and scale separation.
\end{itemize}

Collectively, these contributions enhance the interpretability and reliability of spectral graph analyses applied to subsampled neural data, offering paths towards a robust algorithm for inferring community structures from neural data.

\newpage

% In the point process formulation, the probability that a neuron fires a spike depends on the instantaneous difference between the membrane voltage and threshold, $p_{\text {fire}} = 1 - \exp [V(t) -V_{\text th}]$.

% Generalized linear models 
% point and count process models describe the likelihood of 

% \subsection*{Escape rates, spike-response, generalized linear models, and linear-nonlinear Poisson}

% % \subsection{Generalized Linear Models (GLM)}
% Convex optimization problem for exponential function.

\newpage
\section*{Appendix: Leaky integrate-and-fire models}
Leaky integrate-and-fire (LIF) neuron models are a simplification of biological neurons as RC circuits: the membrane voltage $V_m(t)$ decays exponentially to a resting potential $V_{L}$ according to a time constant $\tau=RC_m$.  The time-volution of the depolarization $V(t) = V_m(t)-V_L$ is given by the ordinary differential equation 
$$
\tau\dot V(t) = -V(t) + RI(t).
$$
where $I(t)$ gives the external driving current.
If the voltage reaches a threshold voltage $\theta$ the neuron emits a spike and the membrane voltage is reset to $V_{\text{reset}}$ for an absolute refractory period $\tau_r$.

Free solutions, When $I(t)=0$ and $V(t=0)=V_0$, are given by exponential decay
$$
V(t) = e^{-t/\tau},
$$
as expected for an RC circuit model. The voltage response to an input current $I(t) = q \delta(t-t_0)$ at time $t_0$ and delta function $\delta$ is
$$
V(t) = q e^{(t_0-t)/\tau} \Theta(t-t_0)/C_m 
$$
where $\Theta(\cdot)$ is the Heaviside step function.

\subsection*{Clustered networks of LIF neurons}
Clustered networks of LIF neurons were numerically simulated with balanced network parameters, as described previously~\citep{Rostami2024}. Each cluster of neurons contains excitatory and inhibitory neurons in proportions that are representative of empirical observations of the cortical circuits. Individual neuronal dynamics evolve according to the ordinary differential equation
$$
\dot V_i(t)=\frac{-\left[V_i(t)-V_L\right]}{\tau}+\frac{I(t)+I_{\mathrm{syn}}(t)}{C_m}
$$
where $I(t)$ is an externally injected current and $I_{\mathrm{syn}}(t)$ is the synaptic current input which evolves according to
$$
\tau_{\mathrm{syn}} \frac{d I_{\mathrm{syn}}^i}{d t}=-I_{\mathrm{syn}}^i+\sum_j J_{i j} \sum_k \delta\left(t-t_k^j\right)
$$
where $t_k^j$ is the arrival time of the $k$th spike from presynaptic neuron $j$ and $\delta$ is the Dirac delta distribution.

\subsection*{Fully-connected network LIF model}
Consider a network of LIF neurons~\citep{Gerstner1996}. 
$$
\begin{aligned}
& \dot{V}_i=-\frac{V_i}{\tau}+I_i(t) \\
& I_i(t)=\sum_j J_{i j} \alpha\left(t-t_j^f\right)
\end{aligned}
$$
where $J_{ij}$ gives the strength of synaptic coupling and the function $\alpha (s)$ is the post-synaptic current response caused by the pre-synaptic spike. Two simple choices for the post-synaptic current response include $\alpha(s) = \delta(s)$ and $\alpha(s)=\delta(s-\Delta)$, where $\delta$ is the Dirac delta function and $\Delta$ is a delay. If $V_i(t)=\theta$, then a neuron fires a spike at a time denoted by $t_i^f)$ and the state of the neuron is reset according to
$$
\lim_{\delta \to 0} V(t_i^f + \delta) = 0.
$$
The reset is equivalent to an injected current $-\theta\delta(t-t_i^f)$

Integrating the linear ordinary differential equation gives the solution
\begin{align*}
    V_i(t)=&\sum_f \eta(t-t_i^f)+\sum_j J_{i j} \sum_f \varepsilon(t-t_j^f)\\
    % =&\int_0^{\infty} \eta(s) \sum_f \delta(t-t^f_i-s) ds + \sum_j J_{i j} \int_0^{\infty} \varepsilon(s) \sum_f \delta(t-t^f_j-s) ds\\
    =& \int_0^{\infty} \eta(s) S_i(t-s)ds+\sum_j J_{i j} \int_0^{\infty} \varepsilon(s) S_j(t-s)ds
\end{align*}
with spike train $S_i = \sum_f \delta(t-t^f_i)$ and autoregressive convolutional filters\footnote{Linear convolution: $(f * g)(t) \equiv \int_{0}^{\infty} f(s) g(t-s) d s$.} corresponding to an adaptive spike after-effect or refractory kernel 
$$
\eta(s)=-\theta e^{-s/\tau}
$$
which enables bursting and network coupling filter
$$
\epsilon(s)=\int_0^s \alpha\left(s'\right) e^{-\left(s-s'\right) / \tau} d s'.
$$

\section*{Appendix: Generalized linear models}
Generalized linear models are state-of-the-art in computational neuroscience. One may derive generalized linear models from integrate-and-fire models via stochastic point processes. Alternatively, using the firing rate model of a neuron, one may model temporally coarse-grained or binned spike trains via stochastic count processes: these are called Linear-Nonlinear Poisson models. 

\subsection*{Point process theory for diffusion in integrate-and fire model}
Langevin diffusion dynamics in fluctuation-driven LIF model are given by the stochastic differential equation
$$
dV(t)=-\frac{1}{\tau}\left [V(t)+RI(t) \right]dt + \eta(t) dW_t
$$
where $W_T$ is a one-dimensional standard Wiener process. Fluctuations drive the neuron to spike, even if the deterministic trajectory is below threshold, and lead to a diffusion in firing times and interspike intervals. These effects are described within the mathematical framework of stochastic point processes~\citep{Truccolo2005,Gerstner2014}.

% For constant input, the stochastic intensity is the hazard function of the renewal model. We account for time-dependence: basis for the likelihood of a spike train and modern methods for fitting neuron models. 
We assign a probability of the neuron's membrane voltage to cross the threshold even before it has reached the threshold. This instantaneous rate of crossing the threshold, i.e. firing a spike, depends on the momentary difference between the membrane potential $u(t)$ and the threshold $\theta$. This is the stochastic intensity of a point process,
$$
\rho(t)=f\left[V(t)-\theta\right].
$$
One possibility is to set the escape rate as an exponential function.
$$
\rho(t)=\rho_0 \exp \left(\frac{V(t)-\theta}{\beta}\right)
$$
where $\beta$ is inversely related to the steepness.

The survivor function gives the probability that a neuron remains silent from the last spike firing time $\hat t$. It is assumed that the survival probabilities decay exponentially with a time-dependent rate.
$$
\dot{S}(t | \hat{t})=-\rho(t) S_I(t \mid \hat{t}),
$$
Integrating over time gives the survivor function
$$
S(t | \hat{t})=\exp \left[-\int_{\hat{t}}^t \rho\left(t^{\prime}\right) d t^{\prime}\right]
$$
Consider a discrete finite time interval $[0,T]$ where time points are denoted by $t_k$ separated by time intervals $\Delta = t_{k+1}-t_k$ and indexed by $k=0\ldots K$, where $K$ is the number of time points. Times points at which a neuron fires a spike are denoted as $t^f$, indexed by $f=1\ldots N_{1}$, where $N_{1}$ is the number of time points at which a neuron fires a spike. Times points at which a neuron is silent is denoted as $t^s$, indexed by $s=1\ldots N_{0}$, where $N_{0}$ is the number of time points at which a neuron is silent.  The spike train is a time-series for the neuron's spiking activity is represented as a binary vector $x(t)\in [0,1]$.

% \subsubsection*{Likelihood of a spike train}
The probability of a neuron to remain silent in one discrete interval $\Delta = t_{k+1} - t_k$ is given by the survivor function
$$
p(t^s) = S\left(t_{k+1} | t_k\right)=\exp \left(-\int_k^{k+1} \rho\left(t^{\prime}\right) d t^{\prime}\right) \approx \exp \left(-\rho\left(t^s\right) \Delta\right)
$$
where it is assumed that the resolution of the discrete time steps are small enough to enable very small changes in the the survivor probability. The probability of a neuron to fire a spike is
$$
p(t^f) = 1-S\left(t_{k+1} | t_k\right)=1-\exp({-\rho\left(t^f\right)})
$$
which is bounded under 1 since the exponential explodes. 

\subsubsection*{Likelihood of a spike train}
The likelihood of a spike train with firing times at $t^1, t^2,\ldots,t^{N_1}$
$$
L\left(t^1, \ldots, t^{N_1}\right)=\prod_f^{N_1} p(t^f) \cdot \exp\left(-\sum_s^{N_0} p(t^s) \Delta\right)
$$
The log-likelihood is
\begin{align*}
\log L\left(t^1, \ldots, t^{N_1}\right) &=\sum_f^{N_1}\;\log p(t^f) - \sum_s^{N_0}p(t^s) \\
&=\sum_f \log(1-\exp(-\rho(t^f)\Delta) - \sum_s \rho(t^s)\Delta
\end{align*}

In continuous time limit, $\Delta \rightarrow 0$, the spike train for neuron is a sum of delta functions $x(t)=\sum_f \delta\left(t-t^f\right)$. The probability of firing a spike is
$$
p(t^f)=1-e^{-\rho\left(t^f\right) \Delta} \approx \rho(t^f) \Delta
$$
is where Taylor expansion $e^x=1+x+\cdots$ has been used. The probability that a neuron is silent in an interspike interval is given by the survival probability %$S(t^f|t^{f-1})$
$$
S(t^f|t^{f-1}) = \exp \left(-\int_{t^{f}}^{t^{f + 1}} \rho\left(t^{\prime}\right) d t^{\prime}\right)
$$
The discrete-time likelihood of a spike train with firing times at $t^1, t^2,\ldots,t^{N_1}$
\begin{align*}
    L_\Delta\left(t^1, \ldots, t^N\right)
    &= S(t^1|t_0)p(t^1)S(t^2|t^1)p(t^2)\ldots S(t^{N_1}|t^{N_1-1})p(t^{N_1})\\
    &=\Delta^{N_1}\prod_f \rho\left(t^f\right) \exp \left(-\int_0^T \rho\left(t^{\prime}\right) d t^{\prime}\right) 
\end{align*}
is related by an $N_1$-fold integration of the continuous-time likelihood density
$$
L\left(t^1, \ldots, t^N\right)=\prod_f \rho\left(t^f\right) \exp \left(-\int_0^T \rho\left(t^{\prime}\right) d t^{\prime}\right).
$$
The log-likelihood density in continuous time is
$$
\log L\left(t^1, \ldots, t^N\right)=\sum_f^{N_1} \log \rho\left(t^{f}\right)-\int_0^T p\left(t^{\prime}\right) d t^{\prime}
$$
The integrate-and-fire model with a stochastic intensity $\rho(t)$ can be seen as a generative model: if we observe a spike train, we can calculate the likelihood that it could have been generated by the model. The model has parameters that one can optimize.

% We also have a time-dependent or dynamic threshold
% \begin{align*}
% \theta(t)&=\theta_0+\int_0^{\infty} \theta_1(s) S(t-s) d s\\
% &=\theta_0+\sum_f \theta_1(t-t^f)
% \end{align*}

\subsection*{Count process Linear-Nonlinear Poisson}
Linear nonlinear Poisson are special case of generalized linear model. However, the LNP model has some limitations~\citep{Pillow2008}. For example, it does not account for factors like refractory periods or other history-dependent aspects of neuronal spiking. Spike history filters can be incorporated into GLMs to yield more accurate results. For recordings from a large population of neurons, one can include connections between neurons in the GLM through coupling filters.

Linear-nonlinear Poisson (LNP) cascades model neural spiking as a count process, where the probability of observing $k$ spikes in a discrete time interval $\Delta$ is given by
$$
P(k)=\frac{(\Delta \lambda)^k}{k!} e^{-\Delta \lambda}
$$
where the firing rate $\lambda$ is a nonlinear function of a linear filter $\mathbf{k}$ operating on an observed spike train $\mathbf{x}$  : $f(\mathbf{k} \cdot \mathbf{x})$. In practice, convolutional filters are constructed from a basis of raised cosine functions~\citep{Tseng2022,Pillow2008}.

Poisson distribution
$$
P\left(y_t \mid \mathbf{x}_{\mathbf{t}}, \theta\right)=\frac{\lambda_t^{y_t} \exp \left(-\lambda_t\right)}{y_{t}!}
$$
with rate $\lambda_t=\exp \left(\mathbf{x}_{\mathrm{t}}^{\top} \theta\right)$.

Log likelihood
$$
\log P(\mathbf{y} \mid X, \theta)=\sum_t\left(y_t \log \left(\lambda_t\right)-\lambda_t-\log \left(y_{t}!\right)\right)
$$
matrix notation
$$
\mathbf{y}^{\prime} \log (\lambda)-1^{\prime} \lambda, \text { with rate } \lambda=\exp (\mathbf{X} \theta)
$$

Basis functions parameterized by raised cosine bumps
$$
b_i(x)=\left\{\begin{array}{l}
\frac{1}{2} \cos \left(\frac{2 \pi\left(x-c_i\right)}{w}\right)+\frac{1}{2}, \text { for }\left|x-c_i\right|<\frac{w}{2} \\
0, \text { otherwise }
\end{array}\right.
$$

\subsection*{Convexity of optimization landscape}
Predicting spike times is nonlinear in the parameters but has a convex loss if and only if exponential activation function, predicting voltages is linear in parameters and quadratic loss~\citep{Paninski2004}.

\section*{Appendix: Langevin diffusion in fluctuation-driven neural circuits}

\begin{figure}
    \centering
	\includegraphics[width=0.625\textwidth]{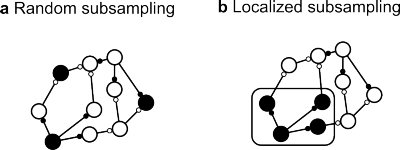}
	\caption{\textit{Partial observation of neural circuits implies different noise models}. \textbf{a}. Random subsampling of sparse neural networks yields homogeneous noise. \textbf{b}. Localized subsampling yields inhomogeneous noise.}
 \label{fig:diss-1}
\end{figure}

Different classes of spatial subsampling yield different noise models. Random subsampling of sparse neural networks implies homogeneous Gaussian noise~\citep{Brunel2000} and localized subsampling allows for non-Gaussian noise (Fig.~\ref{fig:diss-1}). One may explore these implications and provide generalizations to applications with more experimentally-relevant observations, where localized subsampling leads to non-Gaussian noise.

Langevin diffusion in a network of LIF neurons may be described by a stochastic differential equation of the form,
\begin{equation} \label{eqn:ito-sde}
    d\mathbf{x} =\mathbf{f}(\mathbf{x},t)dt+\mathbf{G}(\mathbf{x},t) d\mathbf{w},
\end{equation}
where $\mathbf{x} \in \mathbb{R}^N$ represents the state of the neural system, $\mathbf{f}(\cdot, t): \mathbb{R}^N \rightarrow \mathbb{R}^N$ is a drift term, $\mathbf{G}(\cdot, t): \mathbb{R}^N \rightarrow \mathbb{R}^{N \times M}$ is a diffusion term, and $\mathbf{w}$ is a $M$-dimensional standard Wiener process. 

Langevin dynamics describe a combination of deterministic forces in the drift coefficient $\mathbf{f}(\mathbf{x},t)$, i.e. right-hand side of the LIF model, and stochastic noise described by the diffusion coefficients $\mathbf{G}(\mathbf{x},t)$.
 
Note the general form of the diffusion term allows for different types of noise arising from subsampled observations. For instance, random subsampling of sparse neural networks gives rise to state-independent, homogeneous noise $\mathbf{G}(t)$ whereas biased subsampling of the same networks gives rise to state-dependent, inhomogeneous $\mathbf{G}(\mathbf{x},t)$ noise.

\chapter{Diffusion Tensor Network Renormalization}\label{chapter:dtnr}
 % via Probabilistic Tensors
% \chapter{Probabilistic Tensor Representation Learning}
% \chapter{Renormalizing Probabilistic Models of Neural Circuits}\label{chapter:dtnr}
 % via Multiscale Tensor Networks
We develop coarse-graining transformations for probabilistic models of neural circuits. By examining renormalization group techniques based on pairwise correlation, graph spectral embedding, mutual information, neural networks, Boltzmann machines, and tensor network we identify an overarching problem: while it is possible to describe multiscale flows of neural activity or neural metagraphs with these techniques, it is currently impossible to follow the joint non-equilibrium steady-state distribution of variables at each step of coarse graining using only a finite set of samples. 

Here we enable a multiscale flow in the joint distribution of non-equilibrium neural circuits with variational techniques inspired by tensor networks, quantum circuits, and non-equilibrium thermodynamics. By modeling the non-equilibrium steady-state distribution with an energy-like potential function, we show that one may generate multiscale estimates for collective properties for non-equilibrium thermodynamic systems such as total entropy, attractor dimension, free energy, and heat dissipation. We show that across different hyperparameters of the multiscale model, a selection of critical  values yields multiscale flows that converge to scale-invariant forms. We achieve minimal-complexity models by imposing weight sharing constraints across layers, which reflect permutation and scaling symmetries.

% Developing coarse-graining transformations

% Coarse-graining large neural systems across multiple scales is an open challenge in neuroscience. Recently, renormalization group techniques from statistical physics have revealed scaling properties and fixed points

% enormalization group for neural systems
% Coarse-graining large neural systems across multiple scales is an open challenge in neuroscience. 
% A recent application of renormalization group techniques, inspired by statistical physics, to large-scale neural recording data suggests revealed scaling properties and non-trivial probability distributions, suggesting the collective behavior of neural activity is described by a non-trivial fixed point. 
% While these correlation and PCA-based renormalization group techniques reveal important scaling and collective properties of neural systems,

% to revealing scaling properties and 

% We propose a variational renormalization via tensor networks.

% Exploring renormalization group techniques for neural circuits has recently been identified as a keystone problem in systems neuroscience. In this chapter, several renormalization group techniques for stochastic neural circuits explored. Three classes are tested on numerical simulations and experimental data: correlation and PCA-based coarse-graining, diffusion-based Lapacian renormalization, and dissipative tensor network renormalization.

\section{Introduction}
% \subsection{Coarse-graining neural activity from large-scale recordings}
Inferring the collective properties of complex systems from partial observations is a major challenge, which is particularly severe in systems neuroscience, called the subsampling problem~\citep{Levina2022}. One promising approach leverages renormalization group techniques from statistical physics to identify scale-invariant properties by iteratively coarse-graining subsampled neural population activity~\citep{Meshulam2019}.

\cite{Meshulam2019} introduced a coarse-graining procedure for neural activity based on the covariance structure of interacting neurons. Briefly, they used a greedy algorithm to coarse-grain neural activity in real-space: in each iteration, the activity of the most correlated pairs of neurons was summed together. Further, they argued that principal components analysis  realized a momentum-space coarse-graining procedure, based on the fact that the Fourier transform diagonalizes the covariance matrix in a system with translational symmetry.

These covariance-based procedures helped reveal scaling properties in stationary and dynamic variables, and flows to fixed non-Gaussian distribution of coarse-grained variables. However, in the authors point to a \textbf{key problem} in dealing with real data:

\begin{quote}
We would like to follow the joint distribution of variables at each step of coarse graining, but this is impossible using  only a finite set of samples...When we use the renormalization group to study models, we
 indeed follow the flow of the joint distribution in various
 approximations. When we are trying to analyze data,
 either from experiments or from simulations, this is not
 possible.
\end{quote}

Given the content of this statement, we propose coarse-graining a generative model of neural activity. Specifically, here we construct a modular pipeline from large-scale neural recording data to generalized linear models and multiscale tensor networks. While one might also employ variational renormalization group techniques based on Ising spin-glass models, Boltzmann machines, and neural networks~\citep{Mehta2014}, our particular choices are motivated by capturing the non-equilibrium dynamics of dissipative neural circuits.

% variational renormalization group approach to models.

% here we propose a model-based approach to coarse-grain

% Our first observation is that translational symmetries implicitly assume the underlying geometric domain is a grid; however, the spatial extent of network coupling between neurons is more accurately constrained by permutation symmetries of a graph. 

 % Multiscale tensor network decomposition of a probabilistic model
 % of non-equilibrium neural circuits
% Our goal is to enable a multiscale renormalization group flow in a space of probabilistic models by efficiently modeling the non-equilibrium steady-state probabilities of a neural circuit. 
% We are interested in building a probabilistic model of neural states. We want a computationally tractable probabilistic model of non-equilibrium steady-state probabilities $p(\mathbf{x})$ of neural circuit states $\mathbf{x}$, where the time-dependence has been dropped.

% We are interested in connecting the steady-state joint probability of a binary codeword state $\mathbf{x}$ across the network $p(\mathbf{x})$ to the marginal probabilities for a individual neuronal spiking $p(x_i)$, which can be inferred from large-scale neural recording data via generalized linear models. Representing the joint probabilities may be achieved with a higher-order tensor, but the number of entries would grow exponentially with the number of neurons. 

\newpage
\section{Results}
\subsection{Probabilistic tensor network representation of joint distributions}
We consider a multivariate random variable represented by the vector $\mathbf X = [X_1,...,X_N]$ for $N$ discrete random variables taking values in $1,\ldots,d$. \cite{Glasser2019} argued the joint probability function $p(\mathbf X)$. In general, this joint probability function may be resented as a tensor with $N$ indices, each of which can take $d$ values. For each configuration $\mathbf{X}$, the tensor element $T_\mathbf{X}$ stores the probability $P(\mathbf{X})$. When $N$ is large, the number of tensor elements scales exponentially with $N$. It is impossible to store $T$. When there is some structure in the variables, one may exploit the structure to build a compact representation of $T$ with probabilistic graphical models, such as Bayesian networks or Markov random fields. Here we consider models known as tensor networks, in which a tensor is decomposed into the contraction of many smaller tensors.

\begin{figure*}[h]
  \centering
  \includegraphics[width=\textwidth]{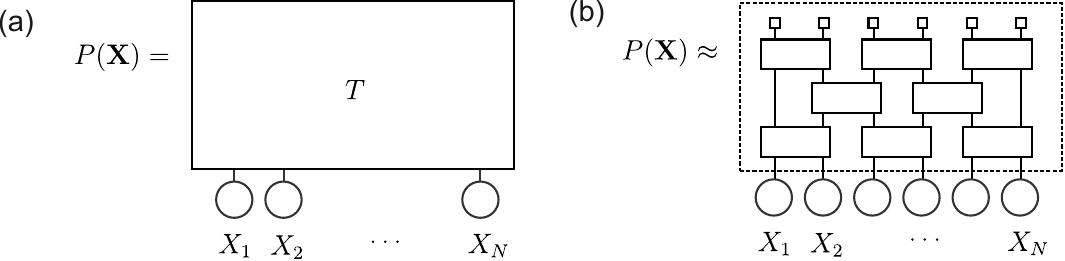}
  \caption{Probabilistic tensor representation with a multiscale tensor network decomposition. (a) Tensor representation of the joint probability for a collection of random variables. (b) Approximation with a tensor network decomposition.}
  \label{fig:tensor-decomposition}
\end{figure*}

Efficient tensor decompositions overcome the curse of dimensionality by sparsely connecting lower-dimensional factor tensor operations to variables of interest. In many-body physics, these tensor network decompositions were motivated by finding ground state wavefunctions via variational renormalization~\citep{White1992,Vidal2007,Evenbly2015}. Recently, tensor networks have been connected to classical data-driven applications~\citep{Stoudenmire2018,Stoudenmire2016}, deep neural networks~\citep{Levine2019}, and probabilistic graphical models~\citep{Glasser2019}.
\begin{figure*}[h]
  \centering
  \includegraphics[width=0.95\textwidth]{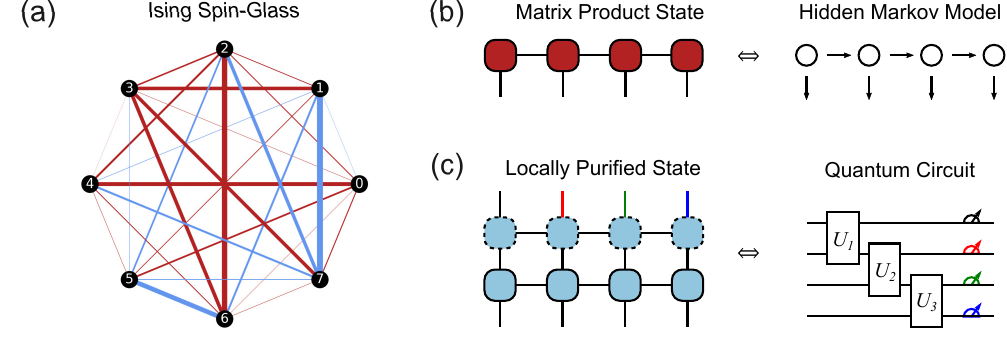}
  \caption{Modeling an Ising spin-glass with MPS and locally purified states. MPS tensor networks map to hidden Markov models and locally purified states map to quantum circuits. \textbf{a}, Ising spin-glass models have symmetric coupling, energy functions, and equilibrium distributions. \textbf{b}, Real-valued MPS tensor networks are equivalent to a hidden Markov model. \textbf{c}, Complex-valued locally purified states are more expressive and correspond to the partial trace over a quantum circuit where measurements are deferred except one.}
  \label{fig:ising-mps-models}
\end{figure*}
Correspondences between probabilistic graphical models and tensor networks allows one to represent probabilistic graphical models as tensor networks via factor graphs: e.g., hidden Markov models are naturally represented as matrix product states (MPS). This implies that $\epsilon$-machines may also be represented by MPS tensor networks, which was discovered independently~\citep{Yang2018}. 

Further, \cite{Glasser2019} demonstrated that these models are just as expressive as Born machines, which are naturally related to the probabilistic interpretation of quantum circuits. Put another way, probabilistic graphical models map to tensor networks and tensor networks map to quantum circuits. Expanding upon the latter, Glasser \textit{et al}. found that quantum locally purified state (LPS) tensor networks wer more expressive than MPS tensor networks; they also studied parameterizations with complex numbers, which led to arbitrarily large reductions in the number of parameters of the networks when compared to parameterizations with real numbers. 
% These results motivate classical simulations and hardware implementations of quantum tensor networks.

Modeling an 8-spin Ising model with MPS and LPS tensor networks demonstrates these differences in expressive power. Convergence times are faster for the LPS tensor network. The KL-divergence between the true joint distribution of the Ising model and the inferred joint distribution of the real-values MPS is $0.074$, while that of the LPS is $0.024$. While MPS tensor networks are very efficient~\citep{Novikov2015,Stoudenmire2016}, but we know that correlations will at some point fall off exponentially; in contrast, multiscale tensor networks are capable of modeling long-range correlations that fall off algebraically.

\begin{figure*}[h]
  \centering
  \includegraphics[width=0.95\textwidth]{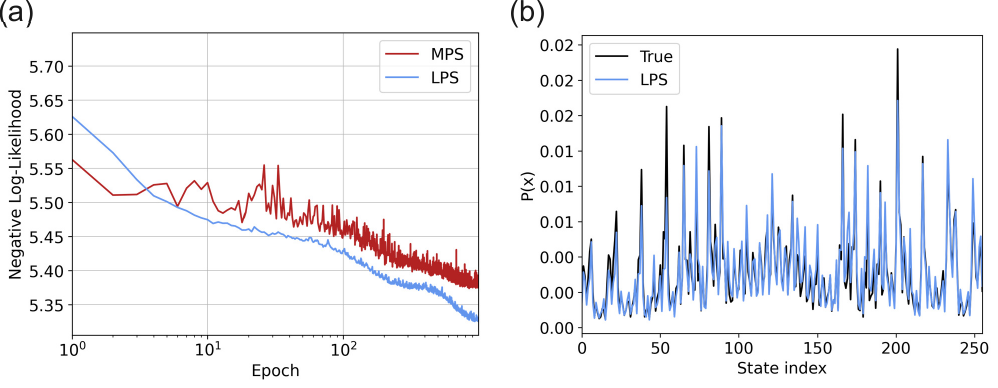}
  \caption{Inferring the joint distribution of an Ising spin-glass with classical MPS and quantum LPS tensor networks. (a) Convergence times are slightly faster for the LPS tensor network compared to that of the MPS tensor network. (b) The joint distribution inferred by a LPS tensor network closely matches the true joint distribution of the Ising spin-glass.}
  \label{fig:ising-mps-results}
\end{figure*}

In the next section, multiscale quantum tensor networks are used to generate a renormalization group flow in the space of joint probability distributions. By isometrically compressing the non-stationary joint distribution of non-equilibrium neural circuits, one can estimate the log-dimension and entropy of the coarse-grained state-space.
% By modeling the non-equilibrium steady-state distribution with an energy-like potential function, we show that one may generate multiscale estimates of total entropy, attractor dimension, free energy, and heat dissipation. 
Searching over hyperparameter values controlling the compression rate of the procedure yields convergence to scale-invariant entropy flows.

\subsection{Multiscale renormalization inspired by quantum tensor networks}
\begin{figure*}[h]
  \centering
  \includegraphics[width=0.75\textwidth]{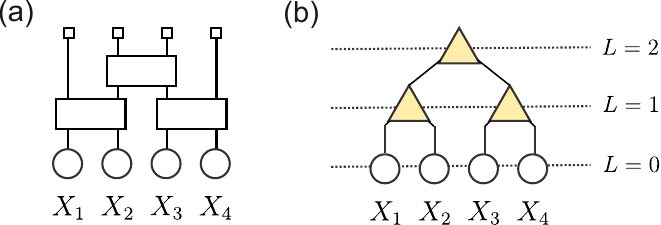}
  \caption{Multiscale renormalization inspired by quantum circuits and many-body systems. (a) Probabilistic circuit representation of a hierarchical tree. (b) Tree tensor network generates a renormalization group flow across multiple scales or levels $L$.}
  \label{fig:ttn-rg}
\end{figure*}

Quantum tensor networks are used to develop a multiscale renormalization procedure for classical neural data. In the probabilistic framework described previously, the procedure is represented by a probabilistic circuit with a multiscale binary tree structure (Fig.~\ref{fig:ttn-rg}a). The tree tensor networks generates a renormalization group flow in the space of joint probability densities across multiple scales or levels $L$ (Fig.~\ref{fig:ttn-rg}b).

Rather than acting directly on Bernoulli random variables, the multiscale tensor network operates on square-integrable states represented by the vector $|p\rangle = [\sqrt{1-p(x_i)},\sqrt{p(x_i)}]^\top$, where we use bra-ket notation to denote vectors. This is inspired by quantum wavefunctions: the mapping to a quantum bit would be exact if the probabilities were encoded in a complex vector $|q\rangle = [\sqrt{1-p(x_i)},i\sqrt{p(x_i)}]^\top$. 

The multiscale renormalization relies on low-rank matrix approximations to density matrices. The state above as a density matrix is given by the outer product $\rho = |p\rangle \langle p|$,
$$
\rho_{i} =\begin{pmatrix} 1-p(x_i) & \sigma_{x_i}\\ \sigma_{x_i} & p(x_i) \end{pmatrix},
$$
where $p(x_i)$ is the probability that a neuron $i$ fires a spike and $\sigma_{x_i}=\sqrt{p(x_i)(1-p(x_i))}$ is the standard deviation. Note that in a purely classical setting, the off-diagonal elements would be zero. It will be useful conceptually to refer back to the classical case as we proceed.
% however, keeping the off-diagonal elements allows for computational advantages offered by coherence.

We assume conditional independence between each neuron's firing probability, $p(x_i)$ and $p(x_j)$: given the state $h_i$ of a neuron, such as its voltage, the probability of firing a spike depends solely on the instantaneous distance to its threshold and not on the state of the rest of the network. Therefore, the entropy for the neuron is given by the trace of the density matrix $S_i = \operatorname{Tr} \rho_i$ and the total entropy of the system is $S_{\text sys} = \sum_i S_i$. Compositions of neurons is given by the Kronecker product of the respective density matrices $\rho_{ij} = \rho_i \otimes \rho_j$. It follows that the joint distribution is $p(x_i,x_j) = \operatorname{diag} \rho_{ij}$.

We are ultimately interested in the renormalization group flow in the space of joint distributions, however, this is obstructed by the curse of dimensionality: the joint distribution grows as $2^N$, where $N$ is the number of neurons. Inspired the density renormalization group in quantum many-body systems, we progressively find lower-rank approximations to compositions of density matrices using singular value decomposition. 

To ensure density matrices are full rank, we construct the separable state,
$$
\rho_{ij} = \sum^M_m \rho_{i}(t_m) \otimes \rho_{i}(t_m),
$$
where we average over a batch of $M$ examples at discrete points in time. Singular value decompositions of the separable state $\rho_{ij} = U\Sigma V^\dagger$ iteratively generate low-rank approximations. Each step of multiscale renormalization procedure proceeds by (1) truncating the singular values according to some error tolerance $\varepsilon$ and (2) projecting density matrices at each time point with the corresponding columns of $U$. Such isometric compression may be interpreted as a change-of-basis or rotation and then projection to a lower-dimensional space. For purely classical density matrices have non-zero elements only on diagonal, so there is no change of basis, there is only an elimination of the lowest probability states in the joint distribution.

\subsection{Renormalization group flows in system size and entropy}
\begin{figure*}[h]
  \centering
  \includegraphics[width=1\textwidth]{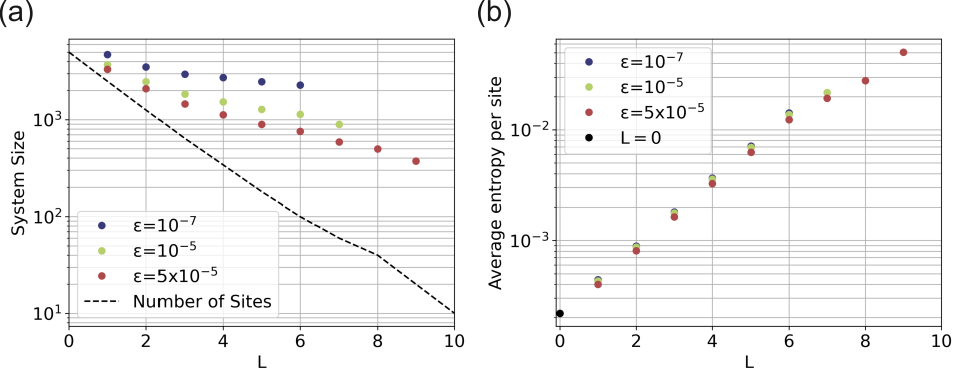}
  \caption{Renormalization group flows in system size and entropy per site across a selection error tolerances. (a) System size decreases exponentially in the number of sites and sub-exponentially in the log-dimension. (b) Average entropy per site increases exponentially and is stable across the selection of error tolerances.}
  \label{fig:isometric-compression}
\end{figure*}

Clustered neural networks, as described in Chapter~\ref{chapter:lgd}, are simulated and renormalized with multiscale tensor networks. At each iteration of the coarse-graining, pairs of neurons are combined according to the procedure above, leading to an exponential reduction in the number of sites and exponential increase in entropy per site (Fig.~\ref{fig:isometric-compression}).

Across a selection of error tolerances, the log-dimensionality $\log_2 \sum_i \operatorname{rank}(\rho_{i,L})$ decreases sub-exponentially (Fig.~\ref{fig:isometric-compression}a), while the time-averaged entropy per site $S_L = \sum_n^{N_L}\langle S_n(t)\rangle_t/N_L$ increases exponentially (Fig.~\ref{fig:isometric-compression}b). The number of effective degrees of freedom reaches a minimum value for $\varepsilon=5\times 10^{-5}$, which corresponds to a $ {10}\times$ compression in system size or, equivalently, a $2^{10}$ reduction in dimensionality. These results suggest that information is being compressed into exponentially smaller numbers of degrees of freedom as the system is coarse-grained.
% described in~\ref{cha}

\begin{figure*}[h]
  \centering
  \includegraphics[width=1\textwidth]{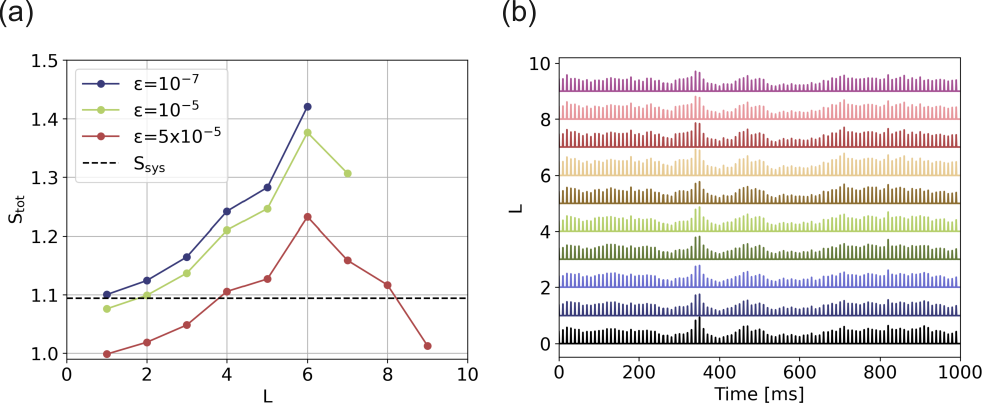}
  \caption{Stable renormalizing flows in total system entropy across error tolerances. (a) Total system entropy is approximately stable to coarse-graining transformations. (b) Dynamic variations in the system entropy are preserved in the renormalization group flow.}
  \label{fig:critical-flows}
\end{figure*}

Total entropy of the system is approximately stable to coarse-graining transformations (Fig.~\ref{fig:critical-flows}a). At fixed error tolerances, the total system entropy grows modestly with each coarse-graining step $L$; however, for larger error tolerances, the trend reverses halfway through the coarse-graining procedure. Entropy dynamics are preserved in the renormalization group flow, i.e. the dynamic variations in the total system entropy are scale-invariant (Fig.~\ref{fig:critical-flows}b).

% \subsection{Robustness to dynamical subsampling}

% \subsection{Inferring latent representations of community-clustered neural networks}
% \subsection{Latent graphs inferred from stochastic graph diffusion models}

% \subsection{Robustness of DGR framework to permutations}
% \subsection{Improvements to entropy calculations}

% \subsection{Improvements to entropy flows with increasing bond dimension}
% predictive embedding and coarse-to-fine reconstruction

% \subsection{Application to coarse-graining large-scale neural data}

\newpage
\section{Discussion}
Coarse-graining connects tensor networks from many-body physics and machine learning to complex networks of neurons. Specifically, the strong disorder MERA and PCA based TTN are tested against the real- and momentum-space RG procedures put forth by~\citep{Meshulam2019}. It is shown that the former will lead to higher-precision estimates of scaling in static and dynamic variables for more RG steps, whereas the latter will offer practical algorithmic speedups in high-dimensional systems with more data. In the case of strong disorder MERA, it appears that it is equipped to overcome a potential breakdown in~\citep{Meshulam2019}'s real-space RG procedure in large systems because it inherits the hierarchical structure of correlations and prevents their propagation to coarser scales via a variational RG transformation. In the case of the PCA based TTN, there is a cubic or quadratic speedup in the dimension of the system taking PCA from polynomial scaling $O(Nd^2+d^3)$ to linear scaling $O(Nd)$. The algorithmic corrections and speedups will likely prove crucial as the work moves towards extending tensor networks to perform transformations necessary for dynamical modeling. In particular, the interest ia in replacing deep learning with tensor networks in nonlinear dimensionality reduction.

The analysis begins by considering coarse-graining networks of neurons separately across space and time---however, there are both practical and fundamental motivations for renormalizing neural systems in both space and time. There are several ways to increase the expressive power of tensor networks. The \textit{lattice dimension} of the tensor network can be increased---this leads to tensors with more indices and pairwise connections---however, this leads to an exponential increase in both the expressive power and computational cost. However, some 2d tensor network models, such as the 2d generalization of matrix product states, the Projected Entanglement Pair States (PEPS), is tractable. If there is symmetry in the system, a more computationally sustainable approach is to model features of scale through tensor network-based renormalization. Consider matrix product state and MERA, where the number of tensors in both tensor networks scales linearly with the number of inputs, but the mean path length grows linearly in a matrix product state and \textit{logarithmically} in MERA. Because pairwise correlations are inversely proportional to path lengths, MERA can capture power law correlations while matrix product states can only capture exponentially decaying correlations.\footnote{Note that the depth of the corresponding quantum circuit, which is a measure of computational complexity, also scales linearly for matrix product states and logarithmically with MERA.} Expressive power may be increased in other ways,\footnote{The internal \textit{bond dimension} between variational tensors may be increased, which leads to a greater number of parameters and more expressive power---computational complexity scales polynomially with the bond dimension. In tandem, consider also enforcing symmetries--periodic boundary conditions decreases the mean path distance between sites and translation invariance decreases the number of tensors to optimize.} but these practical considerations only partially motivate further generalizations of coarse-graining transformations. Fundamentally, coarse-graining yields an RG flow in the space of models across different scales. Non-equilibrium systems that are critical or chaotic do not have a characteristic length scale. Therefore, it is important to investigate features of scale as obtained by a proper RG procedure by describing instances in which MERA achieves this in real-space, separately for space and time, and a case for TTN-based PCA for momentum-space coarse-graining in spatial coordinates.

\newpage
\section{Contributions}
In this chapter, I developed the diffusion tensor network renormalization framework that addresses the fundamental challenge of modeling large-scale neural systems. This method enables coarse-graining of joint probability distributions describing neural dynamics, overcoming computational limitations encountered with traditional approaches. The  contributions of this chapter include:
\begin{itemize}
    \item \textbf{Probabilistic Tensor Network Framework}: Introduced probabilistic tensor networks to efficiently model joint probability functions of neural systems. This framework compresses large-scale neural data, enabling high-fidelity and low-dimensional representations of neural networks.
    \item \textbf{Diffusion Tensor Network Renormalization}: Constructed and demonstrated the diffusion tensor network procedure, inspired by variation quantum circuits, enabling the generation of multiscale renormalization group flows.
    \item \textbf{Multiscale Estimation of Collective Properties}: Enabled the estimation of collective  properties: total entropy and attractor dimension. Total entropy was shown to be locally stable against multiscale transformations, suggesting it is scale-invariant.
\end{itemize}

Together, these contributions establish diffusion tensor network renormalization as a potent approach to scalable and interpretable modeling of non-equilibrium neural dynamics, enabling a deeper understanding of the emergent properties of complex neural systems.

\newpage
\section*{Appendix A: Background on tensor networks}
Tensor network models originated from many-body physics, but they have proved useful in other domains including quantum computing, artificial intelligence, and probabilistic modeling~\citep{Orus2019}. Based on renormalization group (RG) techniques, tensor networks were created to efficiently \textit{coarse grain} many-body systems according to their correlation structure. A variational RG, known as the density matrix renormalization group (DMRG)~\citep{White1992}, led to a tensor network known as the matrix product state (MPS) for 1d spin chains with short-range interactions, while later efforts led to the tree tensor network (TTN) and multiscale entanglement renormalization ans{\"a}tz (MERA) for 1d critical systems with long-range interactions~\citep{Vidal2007}. Although they were originally developed for quantum systems, tensor network models extend to classical partition functions~\citep{Evenbly2015} and machine learning~\citep{Stoudenmire2016}; saliently, there are key correspondences between tensor networks and deep learning~\citep{Levine2017,Levine2018,Levine2019,Cong2019}, quantum circuits~\citep{Huggins2019}, and probabilistic models~\citep{Glasser2019}.

Applying tensor networks to neural systems is motivated by emerging data-driven methods for coarse graining critical states with strong disorder~\citep{Meshulam2019}, inferring nonlinear dynamics from neural spiking~\citep{Pandarinath2018}, and expressively modeling high-dimensional neural codes~\citep{Stringer2019a,Stringer2019b}. Data-driven approaches make compelling predictions on large populations of neurons, however, they often lack computationally sustainability, physical interpretability, or expressive power. We discuss how tensor networks may be used to find algorithmic speedups, mechanistic insight, and probabilistic predictions. Exploring these techniques in tandem with analytical methods from mathematical neuroscience~\citep{Engelken2020} offers multiple paths towards a dynamical theory of neural systems. We describe a potential approach that is physically motivated by RG techniques and computationally sustained by tensor networks.

\subsection*{Tensor diagram notation}
Throughout the text, we use the graphical notation of tensor diagrams to represent tensors and tensor operations. Here we distill the essentials of the tensor diagram notation to understand its usage in the main text. We briefly introduce the graphical notation for the basic building blocks, tensor compositions, and tensor contractions. For a more comprehensive background on tensor diagram notation for statistical modeling of complex physical systems, we refer the reader to several texts on the subject~\citep{Glasser2019,Biamonte2019,Bridgeman2017,Glasser2020}

\begin{figure*}[h]
  \centering
  \includegraphics[width=0.95\textwidth]{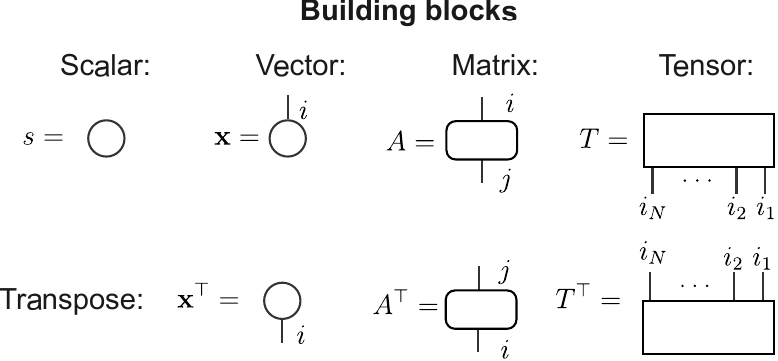}
  \caption{Basic building blocks of tensor diagram notation.}
  \label{fig:building-blocks}
\end{figure*}

Building blocks of tensor diagram notation are illustrated in Figure~\ref{fig:building-blocks}. While tensors are often represented by multidimensional arrays on a computer or algebraic notation in mathematical analysis, they are intuitively represented with nodes and edges in a graph. Each tensor is represented by a node and its indices are represented by edges. The order of a tensor is given by the degree or number edges incident on the corresponding node: scalars are order-0 tensors, vectors are order-1 tensors, and matrices are order-2 tensors. Higher-order tensors are represented by a node with $N$ edges, labeled by the set of indices $i_1, i_2, \ldots, i_N$ running clockwise around the node. Tensor transposes are achieved by flipping the nodes vertically across the horizontal axis, which reverses the indices to running counter-clockwise around the node.

\begin{figure*}[h]
  \centering
  \includegraphics[width=0.95\textwidth]{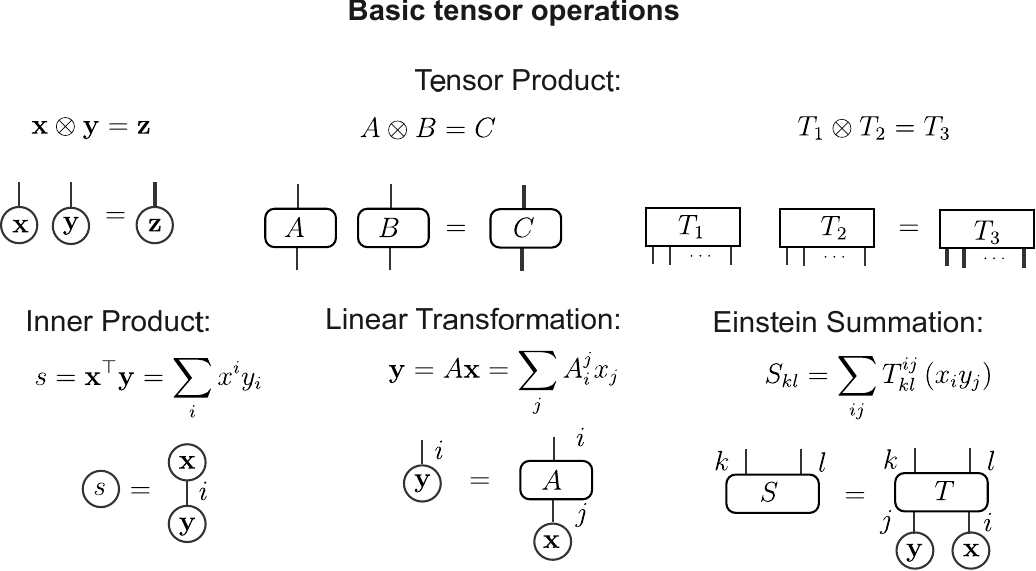}
  \caption{Basic tensor operations for composing tensors and summing over tensor indices. Tensor products compose vectors, matrices, and tensors into higher-dimensional product states. Tensor contractions merging edge-connected nodes represents Einstein summation over the corresponding tensor indices. Inner product between two vectors and matrix-vector multiplication are examples of lower-order tensor contractions.}
  \label{fig:tensor-operations}
\end{figure*}

Basic tensor operations in algebraic and graphical notation are linked in Figure~\ref{fig:tensor-operations}. Tensor products are motivated by composing product states or independent probabilities in probabilistic models of physical systems. Compositions are obtained by placing two tensors side-by-side; the resulting tensor has the same order and higher-dimensional indices, depicted by thicker edges. In general, the dimensionality of the resulting indices is the product of the dimensionality of the input indices. For vectors, the dimensionality is given by $\operatorname{dim}(\mathbf{z})=\operatorname{dim}(\mathbf{x}) \times \operatorname{dim}(\mathbf{y})$.

Einstein summations over many tensor indices demonstrates the advantages of using the graphical notation over the algebraic notation. Diagrammatic representations of the inner product between vectors and matrix-vector multiplication give intuition for the more general Einstein summation (einsum) operation over tensors. Contracting edges corresponding to summed over indices demonstrates how the tensor diagram notation can be used to express more complicated einsum operations and enables the design of tensor network decompositions for probabilistic models.

\begin{figure*}[h]
  \centering
  \includegraphics[width=0.95\textwidth]{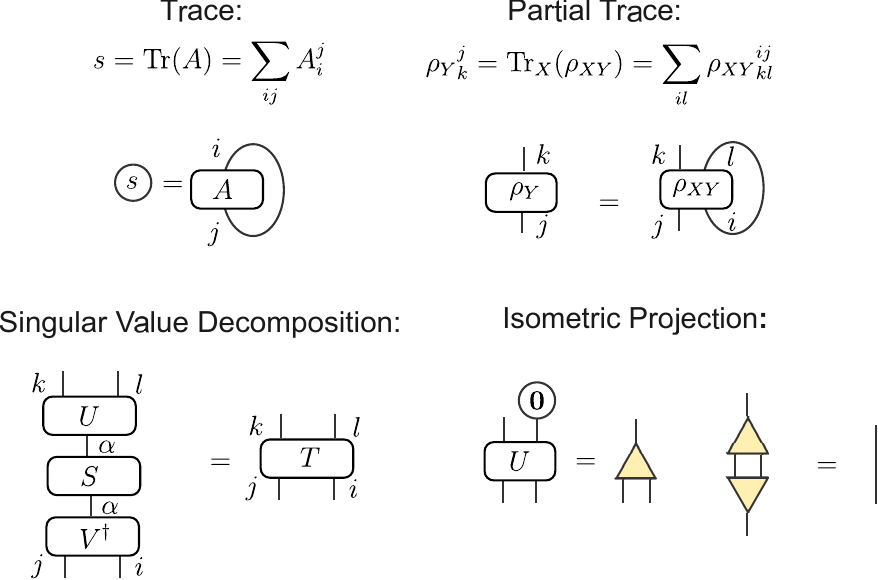}
  \caption{Trace, partial trace, singular value decomposition, and isometry as edge contractions.}
  \label{fig:trace-svd}
\end{figure*}

Trace, partial trace, singular value decomposition, and isometric projection are widely used in tensor network decompositions (Figure~\ref{fig:trace-svd}). For example, calculating observables such as energy and entropy will depend on traces of density matrices, marginal inference from joint distributions and and deferred measurement of subsampled systems will depend on partial traces of composite density matrices. 

Singular value decomposition (SVD) is useful analytically for splitting tensors and numerically for diagonalizing and finding lower-dimensional approximations of tensors in coarse-graining procedures. Isometric projections are obtained by truncating the smallest singular values in $S$ and removing the corresponding subset of basis vectors from $U$. The identity property depicted will be useful in the construction of marginal inference algorithms using tensor network decompositions.

\section*{Appendix B: Coarse-graining neural activity}
Critical states of many-body systems are complex in that they are correlated at every scale. Yet, they are universal: systems with different microscopic details may be mapped to the same universality class by coarse graining. Accurately characterizing the scaling of complex systems with high precision and computational efficiency is crucial to understanding emergent phenomena in these systems. Coarse graining spatially extended networks with nonlocal interactions is a challenging problem in general since traditional RG approaches apply simple rules to combine locally interacting degrees of freedom.

\begin{figure}[h!]
  \centering
  \includegraphics[scale=0.8]{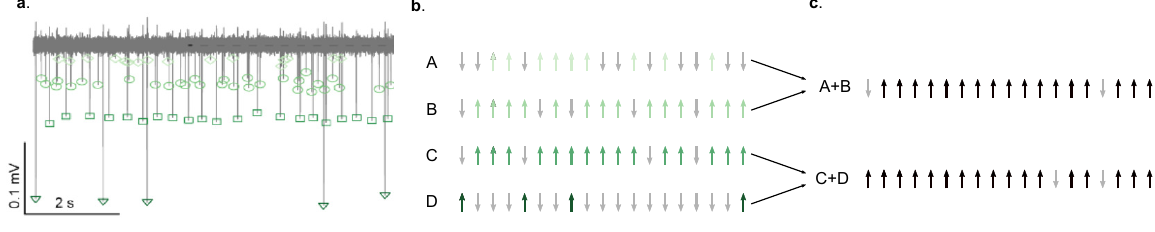}
  \caption{Real-space coarse graining of neural spiking activity. \textbf{a}, Spiking activity of multiple neurons is detected and separated by their features. \textbf{b}, Spins may represent binned spiking activity, where "up" spins corresponds to a spike and "down" spins corresponds to silence. \textbf{c}, Real-space RG via a binary tree in which the temporal activity of strongly correlated neuron pairs is summed.}
  \label{fig:f1}
\end{figure} 

Building upon prior work in this area~\citep{Bradde2017,Tkacik2015}, Meshulam \textit{et al}. (2019) studied real- and momentum-space coarse graining methods in networks of $\sim$1,000 neurons in the hippocampus of freely behaving mice. They discovered features of self-similarity---i.e., power law scaling in static and dynamic variables and distributions that approached non-trivial fixed forms---and thereby provided phenomenological evidence for critical states in the brain. Both their real- and momentum-space RG procedures coarse grained networks of neurons according to their correlation structure. In their real-space RG approach, they successively blocked microscopic degrees of freedom with nonlocal interactions by finding the most correlated pairs and summed their temporal activity (\hyperref[fig:f1]{Fig. 1}). \cite{Meshulam2019} also applied a momentum-space RG approach based on Principal Components Analysis (PCA) in which they eliminated principal components with small eigenvalues (i.e., modes with high momentum). 

\begin{figure}[h!]
  \centering
  \includegraphics[scale=0.8]{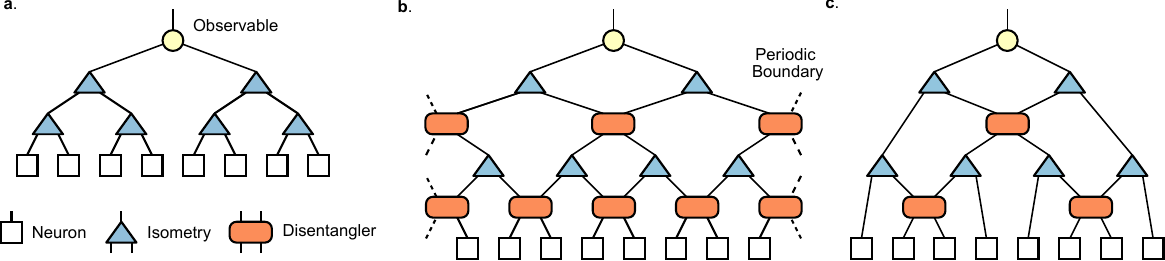}
  \caption{Coarse graining of networks of neurons with tensor networks. \textbf{a}, The PCA based TTN is composed of input vectors corresponding to each neuron (white), isometric transformations (blue), and an observable measurement (yellow). \textbf{b}, The MERA decorrelates degrees of freedom with disentanglers (orange); periodic boundary conditions (dashed) may be enforced. \textbf{c}, Strong disorder MERA combines pairs of degrees of freedom with maximum coupling in a hierarchical tree.}
  \label{fig:f2}
\end{figure}

Meshulam \textit{et al}.’s momentum- and real-space RG procedures serve to port tensor network methods from machine learning and many-body physics, respectively (\hyperref[fig:f2]{Fig. 2}). Their momentum-space RG is closely related to an unsupervised machine learning approach that approximated PCA with a TTN (\hyperref[fig:f2]{Fig. 2a}), which we will refer to as the PCA based TTN~\citep{Stoudenmire2018}. The PCA based TTN offers a key computational advantage: for a system with $d$ dimensions and $N$ examples, PCA scales polynomially $O(Nd^2+d^3)$ whereas the PCA based TTN scales linearly $O(Nd)$---a cubic or quadratic speedup in the dimension of the system. This is a practical algorithmic \textit{speedup} to Meshulam \textit{et al}.'s momentum-space approach, but there is a fundamental algorithmic \textit{correction} to their real-space approach. Their real-space RG procedure is reminiscent of the strong disorder renormalization group (SDRG) from many-body physics, in which pairs of neighboring degrees of freedom with maximum coupling are hierarchically coarse grained ~\citep{Ma1979}. SDRG has since been improved by combining it with a variational ans{\"a}tz known as MERA (\hyperref[fig:f2]{Fig. 2b-c})~\citep{Goldsborough2017}. Crucially, MERA prevents the propagation of microscopic details to coarser scales by removing short-range correlations at each RG step. Both the variational approach to real-space RG and the elimination of short-range correlations were important insights in many-body physics that led to the precise and accurate characterization of critical many-body systems ~\citep{White1992,Vidal2007,Evenbly2015,Evenbly2013}. Thus, both the momentum- and real-space RG procedures of Meshulam \textit{et al}. are extended via tensor networks such as PCA based TTN and strong disorder MERA.

Altogether, the coarse graining techniques employed in Meshulam \textit{et al}. (2019) revealed key scaling features of neural systems. However, their coarse graining procedures do not scale efficiently with the dimension of the system. Further, both the real- and momentum-space procedures are constrained to linear subspaces of RG transformations that may not effectively coarse grain all microscopic details; in fact, the real-space approach propagates short-range correlations to longer time scales, which may lead to mischaracterizations of larger systems. We exhibited two tensor network approaches, TTN based PCA and strong disorder MERA, that address these issues directly. However, there still lacks a clear path to a dynamical theory of neural systems. While others have studied critical neural dynamics~\citep{Cowan2016}, they limit the relevant degrees of freedom to the mean-fields of excitatory and inhibitory populations---this is an oversimplification since neural activity is oftentimes high-dimensional. Data-driven approaches to inferring dynamical models make much more general assumptions and enable the discovery of relevant degrees of freedom and parsimonious equations of motion. 

\section*{Appendix C: Statistical mechanical models}
\subsection*{Equilibrium Ising spin-glass}
Ising spin-glass models are widely used in statistical physics. The state of the system is described by a vector of spins $\vec \sigma$. At equilibrium, the probability of a spin configuration $p(\vec \sigma)$ is given by the Boltzmann distribution
$$
p(\vec \sigma)=\frac{e^{-\beta H(\vec \sigma)}}{Z}
$$
where Z is the partition function
$$
Z=\sum_{\vec \sigma} e^{-\beta H({\vec \sigma})},
$$
$\beta = (k_B T)^{-1}$ is the inverse temperature, with Boltzmann constant $k_B$ and temperature $T$, and $H(\vec \sigma)$ is the energy or Hamiltonian function,
$$
H(\vec \sigma)=-\sum_j h_j \sigma_j-\frac{1}{2}\sum_{i, j} J_{i j} \sigma_i \sigma_j,
$$
with local field strengths $h_j$ and coupling $J_{ij}$.

\subsection*{Non-equilibrium binary spin-glass neural networks with Glauber dynamics}
% The transition probability matrix from coarse-grained state $J$ to state $I$ in an entire sequence of $N$ updates describes a transition taking place coarser time interval: it is the produce of the $N$ transition matrices for the finer time intervals.

% For asynchronous dynamics, the transition matrix depends on the order of updating and not only on the initial and final states. On the coarse time scale, this process is not a Markov chain.
% * Exception: a random sequence, which remains Markovian even on the larger time scale.

% Note that because of the asymmetry, the effective field induced by $J_{ij}$ is not a static field (even at equilibrium) as in the symmetric case.

% \subsubsection*{Numerical simulation}
Binary networks with Glauber dynamics~\citep{Glauber1963} were simulated for networks of size $N_{tot} = 192$, composed of an equal number ($N = 64$) of external (X), excitatory (E), and inhibitory (I) neurons. The networks were updated asynchronously with a single flip update rule. At each time step $dt$, a neuron was selected at random from the population. If the neuron belonged to the external (X) population: a random number $r$ was generated and, if $r < m_X$, the activity of the neuron $\sigma_i^X$ was set to one (otherwise it was set to zero). If the neuron belonged to the recurrent network of excitatory (E) and inhibitory (I) neurons, its afferent synaptic input was calculated using the instantaneous activity of all other neurons in the network that projected on to it; if the resulting current was larger than a threshold $\theta$, its activity $\sigma_i^X$ was set to one (otherwise it was set to zero).

A neuronal time constant of $\tau = 10$ ms was used. The resolution was set to increase with the network size $dt = \tau/N_{tot}$; therefore, each neuron was updated every $N_{tot}$ iterations on average, the neuronal time constant $\tau$. This time constant is how long it takes a neuron to change its activity. Down-sampling the dynamics to $\tau$ then represents a change in the network's activity: this is utilized as a coarse-graining procedure.

The network was simulated for 60,000 $\tau$ (600 seconds). The connectivity of the architecture is as follows: probability of connection $p = 0.2$, mean rate of the external population $m_X = 0.1$, threshold of the recurrent network $\theta = 1$, and synaptic weights $j_{\alpha \beta}$ ($j_{E E} = 5/\sqrt{N}$, $j_{E I} = -10/\sqrt{N}$, $j_{I E} = 5/\sqrt{N}$, $j_{I I} = -9/\sqrt{N}$, $j_{I X} = 4/\sqrt{N}$, and $j_{E X} = 5/\sqrt{N}$.
% \part{Experimental}
\chapter{Towards Coarse-to-Fine Models: Prediction, Control, and Decoding}\label{chapter:c2f}
% \chapter{Coarse-to-Fine Manifold Learning of Dissipative Neural Trajectories}

Understanding how high-dimensional neural trajectories give rise to meaningful computations and behaviors is a central challenge. Dissipative neural dynamics occupy manifold subspaces, exhibiting intricate patterns characterized by complex interactions of stable, neutral, and unstable modes. To unravel these complexities, this chapter develops a generalized framework for coarse-to-fine predictive modeling, decoding, and control of dissipative neural trajectories. Leveraging representation learning, expressive probabilistic tensor networks, and sparse regression techniques, I introduce methods for efficiently modeling and interpreting the underlying structure of neural manifold dynamics. These methods facilitate quantification of crucial dynamical properties---such as stability, entropy, and attractor dimensionality---and enable paths towards linking neural activity to behavioral and information processing.

\section{Introduction}
Dissipative neural trajectories occupy regional subspaces of a high-dimensional phase space by tracing out manifold attractors that are highly non-trivial tangles of stable, neutral, and unstable manifolds~\citep{Engelken2020}. Learning latent dynamical embeddings of neural trajectories is motivated by the prospect of discovering interpretable mappings from neural activity to behavior~\citep{Schneider2023,Liu2022,Batty2019}. Learning the effective dynamics of complex systems is a promising approach to discovering latent dynamical manifold embeddings~\citep{Vlachas2022,Pandarinath2018}. However, high-dimensional learning is obstructed by the curse of dimensionality.  

Geometric priors are crucial to overcoming the curse of dimensionality and learning effective representations of high-dimensional systems~\citep{Bronstein2021}. Designing scale separation priors that are locally stable to deformation is a critical step of inferring effective dynamical models from neural spiking activity. Representational geometry approaches to the effective dynamics of neural systems is an active research area concerned with so-called neural manifolds~\citep{Ehrlich2022,Chung2021,Cohen2020,Chaudhuri2019a,Chung2018}. 

\subsection{Learning the effective non-equilibrium dynamics of attractor manifolds}
Sequential autoencoders have been adapted to generatively model the spiking activity of 200 neurons and accurately predict behavioral variables and reconstruct neural spiking and firing rates by inferring latent factors governed by nonlinear equations~\citep{Pandarinath2018}. However, the authors themselves point to the lack of interpretability in the trained models. In contrast, more interpretable framework uses deep neural networks to find reduced coordinates and sparse least-squares regression to discover nonlinear models from data~\citep{Champion2019,Brunton2016}. It is complementary to work on using deep neural networks to discover coordinates for Koopman operator theory, in which the target coordinate system is linear and therefore amenable to prediction and control~\citep{Lusch2018}.

Deep learning modules may be replaced with tensor networks that are more efficient. Fully-connected layers are the most costly structures in deep neural networks---replacing these with matrix product states has been shown to preserve the expressive power of the layer, while leading to a 200,000$\times$ compression of the dense weight matrix and 7$\times$ overall compression of Very Deep VGG networks~\citep{Novikov2015}. 

Separately, matrix product states and tree tensor networks have been used to construct tensor network representations of convolutional and recurrent neural networks (CNNs \& RNNs), respectively~\citep{Levine2019,Levine2018,Levine2017}, and others have identified commonalities between CNNs and MERA~\citep{Cong2019}. Others have shown that deep learning models for sequence modeling---RNNs, long-term short-term memory (LSTM) units, and GRU---can be outperformed by tensor network representations~\citep{Guo2018,Tjandra2017}.

Imposing scale separation priors from geometric deep learning is a critical step of inferring nonlinear dynamical models from neural spiking activity. The correspondence between deep learning and tensor network models is exploited to extend the latter into this domain, by replacing layers or whole-networks of deep learning models with tensor network structures. Specifically, tree tensor network will be used to represent CNNs and matrix product state will be used to represent fully connected layers or deep recurrent neural networks. Similar to the approach by \citep{Champion2019}, simple and interpretable dynamical models are sought that are mechanistically relevant---to this end, sparse regression techniques are employed to infer nonlinear dynamical models. 

At the level of effective dynamics, we interested in their Lyapunov stability. If effective dynamical models exhibit negative-definite Lyapunov spectra, then they are in a regime of stable chaos, which can be characterized by dynamical flux tubes ~\citep{Monteforte2012,Touzel2019a}. For both integrate-and-fire and firing rate models, their linear stability is characterized by examining eigenspectra of the stability matrix to determine if the system is stable (negative definite), critical/marginally stable (negative semidefinite), or unstable (positive semidefinite). Also, from the Lyapunov spectra of these nonlinear models, their global stability, entropy rate, and attractor dimension are quantified. The interest in these quantities is at the microscopic limit of networks of neurons---coarse-grained networks are characterized throughout the RG flow. Combining tensor network based coarse-graining with systems identification and quantification enables this sort of analysis in large systems. Further, unstable and high-dimensional dynamics in large systems motivate probabilistic approaches to the statistical mechanics of neural ensembles.
\subsection{Neural decoding with multimodal probabilistic models}
% \subsection{Expressive Probabilistic Models for Neural Decoding}
Reducing the dimensionality of non-equilibrium neural systems may lead to irreversible (irreducible) dynamics that are high-dimensional and unstable, i.e. extensive chaos. This motivates studying probabilistic graphical models and the statistical mechanics of neural ensembles in hopes of discovering dissipative structures that are self-organized by spontaneous fluctuations in neural activity. Their associated information architecture may reveal properties of neural information processing~\citep{Walker2020}, network information flows in neural circuits~\citep{Venkatesh2019}, and strategies to infer intrinsic state variables (prediction) and extrinsic environmental variables (decoding)~\citep{Pitkow2017}. Expressive probabilistic models are central to characterizing the statistical mechanics and information architectures of neural systems, because they may lead to better inference and estimates of key information measures. In the previous section, tensor networks and deep learning models were connected. Below, expressive tensor network representations of probabilistic models that scale efficiently on classical and quantum hardware are studied.

\begin{figure}[h!]
  \centering
  \includegraphics[width=\textwidth]{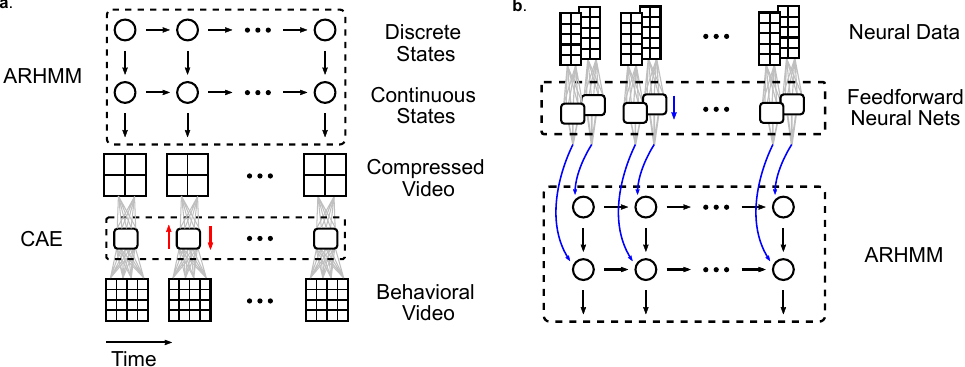}
  \caption{BehaveNet, a probabilistic framework for jointly modeling neural activity and behavioral states~\citep{Batty2019}. \textbf{a}, Generative modeling where CAE compresses behavioral videos; ARHMM segments discrete behavioral states and generates continuous compressed video. \textbf{b}, Bayesian decoder where the ARHMM is treated as the prior, and feedforward neural networks generate conditional distributions over the continuous and discrete states, given a window of neural activity.}
  \label{fig:f5}
\end{figure}

Building on prior sequence modeling approaches employing PCA, VAEs, and probabilistic graphical models~\citep{Wiltschko2015,Johnson2016}, Batty \textit{et al}. (2019) introduce BehaveNet, a probabilistic framework for jointly modeling neural activity and behavioral states of head-fixed mice~\citep{Batty2019}. To better understand the relationship between intrinsic and extrinsic variables, they develop a pipeline for generating full-resolution videos of facial behaviors from neural activity. Using open datasets, they train a convolutional autoencoder (CAE) to compress behavioral videos into continuous latent variables and infer an autoregressive hidden Markov model (ARHMM) that segments the latent dynamics into discrete behavioral states. Treating the behavioral states as a neural decoding target, they construct a Bayesian decoder. Then using the ARHMM as the prior, they train feedforward neural networks to output conditional distributions over the continuous and discrete states, given a window of neural activity. They find that their approach outperforms baseline predictions by $\sim65-75$\%.

Conditioning Batty \textit{et al}.’s decoder on neural activity only leads to a 2\% improvement over the feedforward ARHMM decoder, yielding limited insights into neural coding. Perhaps, better accuracy and insight could be achieved---previously, hidden Markov models (HMMs) have been used to \textit{jointly} model neural spiking and spatial position in mice, effectively coarse-graining two sets of degrees of freedom simultaneously, and accurate reconstructions were decoded with up to $\sim$95\% accuracy~\citep{Box2016}. To gain insight into the statistical mechanics of neural population codes, it is desirable to infer minimal models and characterize their information architecture. For an HMM, this corresponds to an $\epsilon$-machine that groups past states with identical future statistics into equivalence classes known as causal states. In other words, an $\epsilon$-machine uses minimal information from the past to make maximal predictions of the future. Saliently, closed-form expressions may be used to quantify statistical complexity and entropy rate, thereby avoiding sampling errors. \cite{Marzen2020} built upon previous work~\citep{Marzen2017} to model discrete-event processes in continuous-time by inferring a unifilar hidden semi-Markov model (uhsMm) with neural networks~\citep{Marzen2020}. Previously, correspondences between neural networks and tensor networks were described; the relationship between tensor networks and probabilistic graphical models, such as the HMM, is explored below.

\newpage
\section{Methods}
We explore methodologies for learning the effective dynamics of non-equilibrium neural circuits: deep representation learning and energy-based probabilistic models.

\subsection{Deep representation learning of neural populations}
Sequential Variational Autoencoders (VAEs) have been previously adapted to generatively model the spiking activity of $\sim$200 neurons in the motor cortices of rhesus monkeys and humans~\citep{Pandarinath2018}. VAEs accurately predict behavioral variables and reconstruct neural spiking and firing rates by inferring latent factors (hidden degrees of freedom) that are governed by nonlinear equations. This is accomplished by employing bidirectional gated recurrent units (GRUs) to encode single trials of neural activity into a set of initial conditions and generate factors---the latter effectively evolves the dynamics forward in time. They nonlinearly reduce the dimensionality of these factors to visualize manifolds corresponding to different behavioral movements, and show how their model can be trained on single-trials of entirely different populations of neurons, demonstrating the flexibility of the model. 

\begin{figure}[h!]
  \centering
  \includegraphics[width=\textwidth]{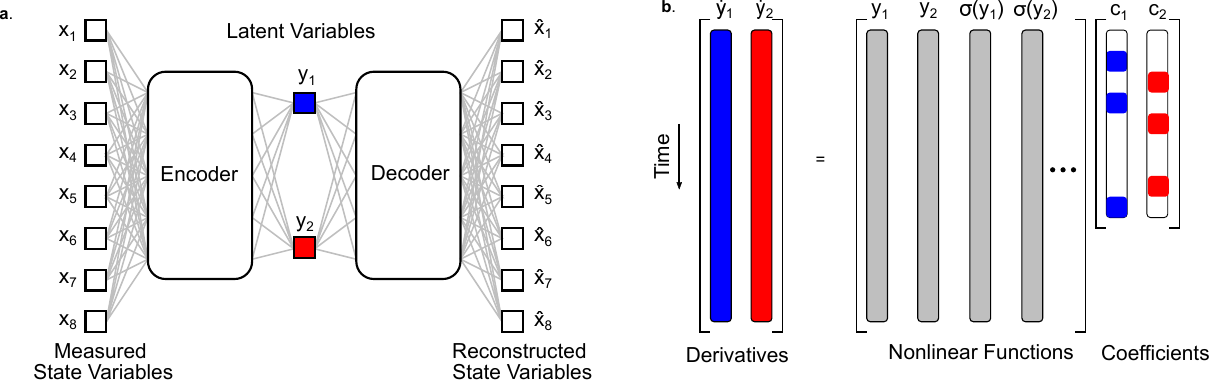}
  \caption{Autoencoders followed by sparse regression~\citep{Champion2019}. \textbf{a}, Autoencoders map measured state variables to latent variables (encoder) such that the reconstruction error is minimized (decoder). \textbf{b}, Assuming derivatives of the data can be computed, sparse regression is used to infer parsimonious models of nonlinear dynamics.}
  \label{fig:autoencoders}
\end{figure}

Although the predictions and visualizations of \citep{Pandarinath2018}’s method are compelling, the authors themselves point to the lack of interpretability in the trained models. In contrast, \citep{Champion2019} built an interpretable modeling framework by proposing a technique for simultaneously discovering coordinates (relevant degrees of freedom) and parsimonious equations of motion. The technique employs deep neural networks to find reduced coordinates and sparse least-squares regression to discover nonlinear models from data (Fig.~\ref{fig:autoencoders}a-b) \citep{Champion2019}. This builds upon their previous work employing sparse regression for model discovery~\citep{Brunton2016}. It is also complementary to their work on using deep neural networks to discover coordinates for Koopman operator theory, in which the target coordinate system is linear and therefore amenable to prediction and control~\citep{Lusch2018}.

\subsubsection*{Representation learning with tensor networks}
Although deep learning models are highly expressive, they may be replaced with tensor networks that are more efficient and interpretable as RG flows. Fully-connected layers are the most costly structures in deep neural networks---replacing these with matrix product states has been shown to preserve the expressive power of the layer, while leading to a 200,000$\times$ compression of the dense weight matrix and 7$\times$ overall compression of Very Deep VGG networks~\citep{Novikov2015}. 

\begin{figure}[h!]
  \centering
  \includegraphics[width=\textwidth]{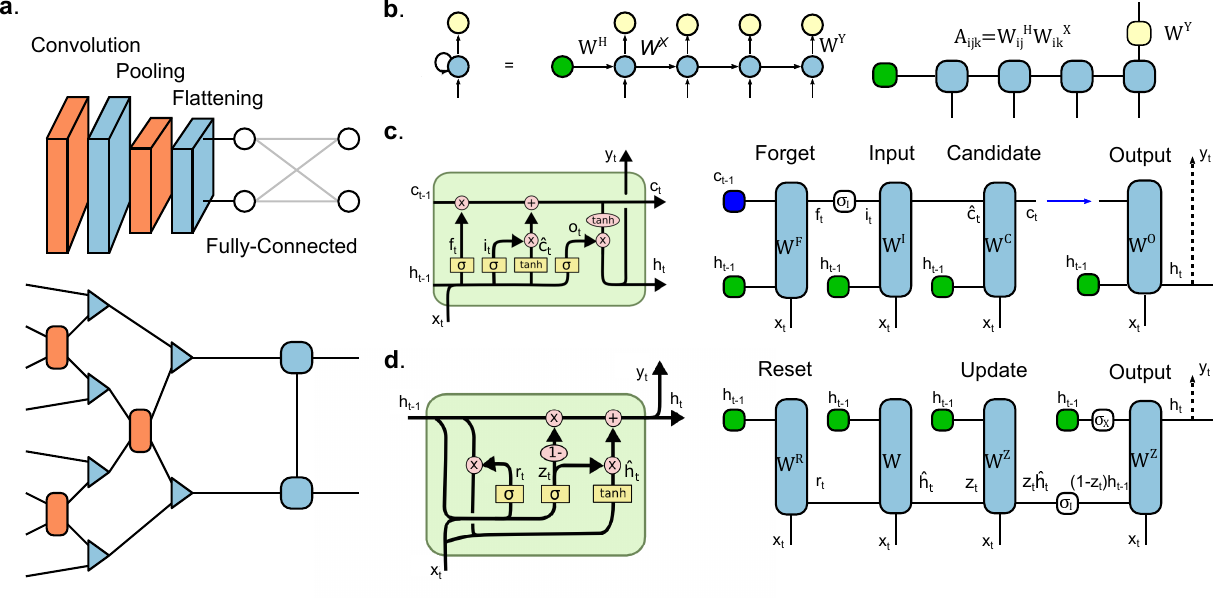}
  \caption{Representation learning with tensor networks. \textbf{a}, CNNs have layers that are closely related MERA: convolutional layers and disentanglers, pooling layers and isometries, and fully-connected layers with matrix product states~\citep{Cong2019, Novikov2015}. \textbf{b}, Simple recurrent neural networks may be implemented with matrix product states~\citep{Levine2019}. \textbf{c}, LSTM has forget, input, gate (candidate), and output gates that can be represented by tensor networks. \textbf{d}, Similarly, GRU has reset, update, and output gates that can be represented by tensor networks.}
  \label{fig:deep-learning}
\end{figure}

Matrix product states and tree tensor networks have been used to construct tensor network representations of convolutional and recurrent neural networks (CNNs \& RNNs), respectively~\citep{Levine2019,Levine2018,Levine2017}, and others have identified commonalities between CNNs and MERA  (Fig.~\ref{fig:deep-learning}a-b)~\citep{Cong2019}. Meanwhile, others have shown that deep learning models for sequence modeling---RNNs, long-term short-term memory (LSTM) units, and GRU---can be outperformed by tensor network representations (Fig.~\ref{fig:deep-learning}c-d)~\citep{Guo2018,Tjandra2017}. Here, an explicit relationship is drawn between tensor networks and sequence modeling by connecting them with probabilistic graphical models. This will serve as another potential path towards a dynamical theory of neural systems via tensor networks.

\begin{figure}[!ht]
  \centering
  \includegraphics[width=\textwidth]{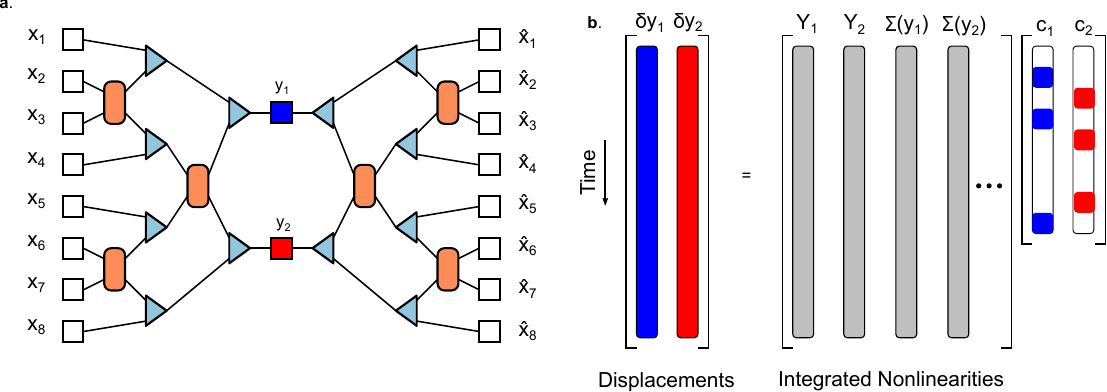}
  \caption{Decomposing representation learning with tensor networks and the integral formulation of sparse regression. \textbf{a}, Strong disorder MERA may replace deep neural networks in autoencoders. Since MERA is a proper RG flow through the space of tensors, I evaluate latent factors and nonlinear models at multiple scales. \textbf{b}, The integral formulation of sparse regression less susceptible to noise. Displacement vectors form the system's initial conditions and integrated nonlinearities are selected by sparse regression.}
  \label{fig:tn-autoencoders}
\end{figure}

Imposing scale separation priors from geometric deep learning~\citep{Bronstein2021} is a critical step of inferring nonlinear dynamical models from neural spiking activity. The correspondence between deep learning and tensor network models is exploited to extend the latter into this domain, by replacing layers or whole-networks of deep learning models with tensor network structures (Fig.~\ref{fig:tn-autoencoders}). Specifically, tree tensor network will be used to represent CNNs and matrix product state will be used to represent fully connected layers or deep recurrent neural networks (Fig.~\ref{fig:tn-autoencoders}a). Similar to the approach by \citep{Champion2019}, simple and interpretable dynamical models are sought that are mechanistically relevant---to this end, sparse regression techniques are employed to infer nonlinear dynamical models. Because the derivative of real data is oftentimes noisy, an integral formulation of sparse regression~\citep{Schaeffer2017} may be considered. This entails calculating displacements and numerical integration of trial functions (Fig.~\ref{fig:tn-autoencoders}b).

\subsubsection*{Quantifying collective dynamical properties}
At the network level, the interest is in fitting integrate-and-fire and firing rate models, and studying their Lyapunov stability. If integrate-and-fire models exhibit negative-definite Lyapunov spectra, then they are in a regime of stable chaos, which can be characterized by dynamical flux tubes~\citep{Monteforte2012,Touzel2019a}. For both integrate-and-fire and firing rate models, their linear stability is characterized by examining eigenspectra of the stability matrix to determine if the system is stable (negative definite), critical/marginally stable (negative semidefinite), or unstable (positive semidefinite). Also, from the Lyapunov spectra of these nonlinear models, their global stability, entropy rate, and attractor dimension are quantified. The interest in these quantities is at the microscopic limit of networks of neurons---coarse-grained networks are characterized throughout the RG flow. Combining tensor network based coarse-graining with systems identification and quantification enables this sort of analysis in large systems. Further, unstable and high-dimensional dynamics in large systems motivate probabilistic approaches to the statistical mechanics of neural ensembles.

\subsection{Energy-based probabilistic model of non-equilibrium neural circuits}
% Our goal is to enable a multiscale renormalization group flow in a space of probabilistic models by efficiently modeling the non-equilibrium steady-state probabilities of a neural circuit. 
We are interested in inferring an energy-based probabilistic model of non-equilibrium neural circuit. We want a computationally tractable energy-based model of non-equilibrium steady-state probabilities $p(\mathbf{x})$ of neural circuit states $\mathbf{x}$, where the time-dependence has been dropped. We define an energy-like potential function $\phi(\mathbf{x})$, which is connected to the non-equilibrium steady-state probability according to
$$
p(\mathbf{x}) = \frac{\exp (-\beta \phi(\mathbf{x}))}{Z}
$$
where $Z = \sum_\mathbf{X} \exp (-\beta \phi(\mathbf{x})$ is the partition function. 

We are interested in connecting the steady-state joint probability of a binary codeword state $\mathbf{x}$ across the network $p(\mathbf{x})$ to the marginal probabilities for a individual neuronal spiking $p(x_i)$, which can be inferred from large-scale neural recording data via generalized linear models. Representing the joint probabilities may be achieved with a higher-order tensor, but the number of entries would grow exponentially with the number of neurons. 

% \subsection{Expressive probabilistic models via tensor networks}
\begin{figure}[h!]
  \centering
  \includegraphics[width=\textwidth]{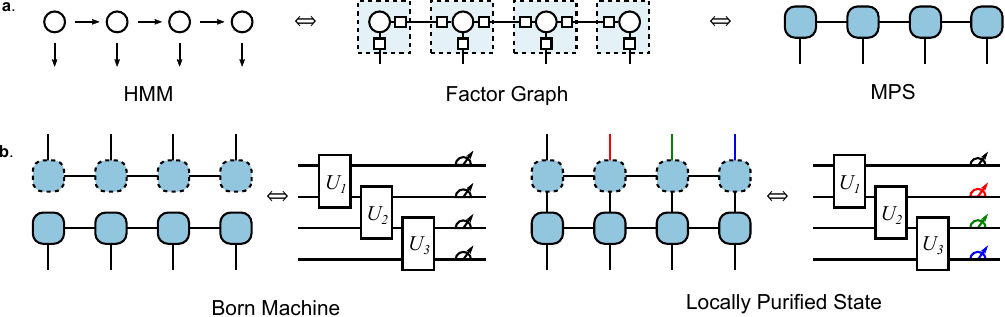}
  \caption{Correspondences between probabilistic graphical models, tensor networks, and quantum circuits~\citep{Glasser2019}. \textbf{a}, An HMM (left) is represented by a factor graph (center) and is equivalent to an MPS (right). \textbf{b}, The Born machine is a product of MPS (solid) with its conjugate transpose (dotted) and maps to the average measurement across a quantum circuit. LPS is a more expressive tensor network and corresponds to the partial trace over a quantum circuit where all but one qubit are unobserved.}
  \label{fig:glasser}
\end{figure}

\cite{Glasser2019} identified correspondences between probabilistic graphical models and tensor networks, and provided a recipe for converting probabilistic graphical models to tensor networks via factor graphs by explicitly connecting HMMs and MPS. This implies that $\epsilon$-machines may also be represented by MPS, which was discovered separately~\citep{Yang2018}. Further, Glasser \textit{et al}. demonstrated that these models are just as expressive as Born machines, which are naturally related to the probabilistic interpretation of quantum circuits. Put another way, probabilistic graphical models map to tensor networks and tensor networks map to quantum circuits. Expanding upon the latter, Glasser \textit{et al}. found that the locally purified state (LPS) from many-body physics was more expressive than MPS; they also studied parameterizations with complex numbers, which led to arbitrarily large reductions in the number of parameters of the networks when compared to parameterizations with real numbers. These results motivate classical simulations and hardware implementations of quantum tensor networks.

Probabilistic graphical models, such as HMMs and ARHMMs, map to tensor network models, as described by Glasser \textit{et al}. (2019). These tensor network representations are used herein to jointly model intrinsic and extrinsic variables. The ARHMM and uhsMm models are combined in a LPS model. This maps neural activity to discrete states that give rise to continuous behavioral variables.  Temporal coarse-graining approaches via the strong disorder MERA may also be considered. This is motivated by recent approaches to using wavelet MERA for regression and classification problems on sequential data~\citep{Reyes2021} and WaveNet, which employs causal, autoregressive CNN to generate audio data~\citep{Oord2016}.\footnote{It is also motivated by computational considerations---tree-like TNs are effective at capturing long-range correlations because the pairwise correlation between two sites is inversely proportional to the mean path length through the tensor network. Considering features of scale, as obtained from MERA, is thus a computationally sustainable approach to increasing the expressive power of the model while also gaining fundamental insight into the system across scales.} Simple and minimal models are found, which in the case of the HMM is an $\epsilon$-machine, with well-defined causal architectures. From these models, statistical complexity and entropy rates of the modeled processes are estimated. The information architecture of neural population codes is of central interest because it may guide further inference on intrinsic state variables (prediction) and extrinsic environmental variables (decoding)~\citep{Pitkow2017}. There is also interest in unrolling the causal architecture in time to reveal information paths. An extensive variety of formulating probabilistic models for neural codes is anticipated, with particular interest in those that are interpretable as multiscale flows.

% \begin{figure}[!ht]
%   \centering
%   \includegraphics[width=\textwidth]{figures/qualifier/f9_computationmech_v2.pdf}
%   \caption{Representations of probabilistic models as tensor network lead to temporal coarse-graining and information architecture. \textbf{a}, An autoregressive tensor network model may be represented as stacked MPS or LPS. \textbf{b}, Strong disorder MERA may be employed as in a temporal coarse-graining procedure that is computationally sustainable to finer time resolution. \textbf{c}, A uhsMm represents discrete and continuous processes with finite-time, e.g. neural spiking. \textbf{d}, The causal states of the uhsMm can be unrolled in time to reveal information paths.}
%   \label{fig:computation}
% \end{figure}

% \section{Results}
% \subsection{Random feature baseline dynamical embeddings of neural trajectories}

% \subsection{Latent dynamics of diffusion-guided tensor network renormalization}
% \subsection{Latent graphs inferred from stochastic graph diffusion models}

% \subsection{Robustness of C2FD framework to dissipation}

% \subsection{Improvements to dynamical predictions with irreversibility}
% % predictive embedding and coarse-to-fine reconstruction

% \subsection{Application to peer prediction and control large-scale neural data}

\newpage
\section{Contributions}
% We outlined an approach to developing expressive probabilistic models for neural decoding by replacing sequence models in an existing joint inference framework with tensor networks. We propose studying open datasets, such as those reported in Batty \textit{et al}. (2019). These include calcium imaging studies of $\sim$10,000 neurons in the primary visual cortex of mice while natural images were presented ~\citep{Stringer2019a} and, separately, electrical recordings of $\sim$3,000 neurons across the brain while facial motion and running speed were tracked~\cite{Stringer2019a}. The pair of studies reported that neural activity had high-dimensional geometry, encoded extrinsic environmental variables, and exhibited correlations that obeyed a power law:  specifically, the variance of principal components fell as a power law. They applied linear decoding analyses and identified the presented stimulus with $89.9$\% accuracy, but did not report any reconstructions of behavioral variables. The high-dimensional geometry of neural codes motivates the application of coarse-graining procedures to these data.

In this chapter, I proposed an approach to coarse-to-fine  modeling, decoding, and control of dissipative neural dynamics. Specifically, the chapter introduces expressive probabilistic tensor network models and sparse regression methods, facilitating interpretable latent embeddings of neural trajectories. The primary contributions include:

\begin{enumerate}
    \item \textbf{Coarse-to-Fine Modeling Framework}: Offered an integrated coarse-to-fine computational framework combining probabilistic tensor network representations and nonlinear sparse regression, aimed at capturing the multiscale dynamics and stability of dissipative neural attractor manifolds.
    \item \textbf{Tensor Networks for Representation Learning}: Introduced tensor-network-based approaches to compressing layers of deep neural networks. Leveraging matrix product states (MPS) and tree tensor networks (TTN), these methods may achieve substantial model compression.
    \item \textbf{Expressive Probabilistic Models for Neural Decoding}: Described multimodal probabilistic tensor network models for joint inference of neural and behavioral states. Illustrated how tensor network-based autoregressive hidden Markov models (ARHMM) and locally purified states (LPS) could enhance decoding performance.
\end{enumerate}

Collectively, these contributions enable scalable and interpretable modeling of dissipative neural dynamics, and offer paths towards the prediction, decoding, and control of complex neural systems.

\begin{singlespace}
\bibliography{main}
\end{singlespace}

\end{document}